\documentclass[pdflatex,sn-mathphys-num]{sn-jnl}

\usepackage{graphicx}%
\usepackage{multirow}%
\usepackage{amsmath,amssymb,amsfonts}%
\usepackage{amsthm}%
\usepackage{mathrsfs}%
\usepackage[title]{appendix}%
\usepackage{xcolor}%
\usepackage{textcomp}%
\usepackage{manyfoot}%
\usepackage{booktabs}%
\usepackage{algorithm}%
\usepackage{algorithmicx}%
\usepackage{algpseudocode}%
\usepackage{listings}%
\usepackage{graphicx}%
\usepackage{multirow}%
\usepackage{amsmath,amssymb,amsfonts}%
\usepackage{amsthm}%
\usepackage{mathrsfs}%
\usepackage[title]{appendix}%
\usepackage{xcolor}%
\usepackage{textcomp}%
\usepackage{manyfoot}%
\usepackage{booktabs}%
\usepackage{algorithm}%
\usepackage{algorithmicx}%
\usepackage{algpseudocode}%
\usepackage{listings}%
\usepackage{enumitem}
\usepackage{graphicx}
\usepackage{booktabs}
\usepackage{amsmath}
\usepackage{booktabs}
\usepackage{graphicx}
\usepackage{booktabs} 
\usepackage[utf8]{inputenc}
\usepackage{enumitem}
\usepackage{xcolor}
\usepackage{textcomp}
\usepackage{csquotes}
\usepackage{xltabular}
 \usepackage{acronym}
\usepackage{graphicx}
\usepackage{hyperref}
\usepackage{rotating}
\usepackage{pgfplots}
\usepackage{multirow}
\usepackage{tikz}
\usepackage{lscape}

 \newacro {CNN}[CNN]{convolutional neural network}
 \newacro {NLP}[NLP]{natural language processing}
 \newacro {NFR}[NFR]{non-functional requirement}
 \newacro {FR}[FR]{functional requirement}
 \newacro {SVM}[SVM]{support vector machine}
 \newacro {CoT}[CoT]{chain of thought}
 \newacro{LLM}{large language model}

\theoremstyle{thmstyleone}%
\theoremstyle{thmstyletwo}%

\theoremstyle{thmstylethree}%

\begin{document}

\title[Article Title]{Are Prompts All You Need? Evaluating Prompt-Based Large Language Models (LLM)s for Software Requirements Classification
\footnote{This preprint has not undergone peer review or any post-submission improvements or corrections. The Version of Record of this article is published in \textit{Requiremente Engineering Journal}, and is available online at \url{https://doi.org/10.1007/s00766-025-00451-8}}
}

\author[]{\fnm{Manal} \sur{Binkhonain}}
\author[]{\fnm{Reem} \sur{Alfayaz}}

\affil[]{\orgdiv{Department of Software Engineering, College of Computer and Information Sciences}, \orgname{King Saud University, P.O. Box 51178},
\orgaddress{\city{Riyadh 11543}, \country{Saudi Arabia}}}



\abstract{Context:
Requirements classification is a fundamental task in software engineering that involves classifying natural language requirements into predefined classes, such as functional and non-functional requirements. Accurate classification is critical to the success of software projects, as it helps reduce development risks and improve overall software quality. However, most existing classification models rely on supervised learning approaches, which demand large, annotated datasets that are resource-intensive, time-consuming, and dependent on domain-specific expertise to acquire. Moreover, these models often struggle with generalizability, as they typically require retraining or fine-tuning for each new classification task.
Objective:
This study aims to address the challenge of dataset scarcity in requirements classification by investigating the effectiveness of prompt-based \acp{LLM}.
Method:
We conducted an empirical benchmarking study to evaluate and statistically compare the performance of multiple prompt-based \acp{LLM} and prompting techniques across several requirements classification tasks on two English requirements classification benchmark datasets: PROMISE and SecReq.
The evaluation incorporated four prompting techniques: zero-shot, few-shot, persona, and \ac{CoT} prompts. For each task, we assessed and statistically compared the performance of different model–prompt configurations. We then compared the best-performing prompt-based \ac{LLM} configurations on each task against the performance of the state-of-the-art fine-tuned transformer-based model.
Results:
The results demonstrate that prompt-based \acp{LLM}, particularly when paired with few-shot prompting, can achieve classification performance comparable to or even exceeding that of the state-of-the-art fine-tuned transformer-based model. Furthermore, augmenting few-shot prompts with either persona alone or a combination of persona and \ac{CoT} can potentially improve model performance.
Conclusion:
This study highlights the potential of prompt-based \acp{LLM} as a practical and scalable alternative to traditional supervised approaches for requirements classification. By reducing the dependency on large annotated datasets, these models provide a flexible solution to overcoming data scarcity and enhancing the generalizability of automated requirements classification tools.}

\keywords{Software Requirements, Software Requirements Classification, Zero-shot Learning, Few-shot Learning, Large Language Models (LLMs)}



\maketitle
\acresetall
\section{Introduction}\label{sec1}

Requirements classification is a fundamental aspect of requirements engineering that involves organizing software requirements into distinct classes. Such classification enhances the understanding, analysis, and management of requirements throughout the software development lifecycle, and it ensures that both functional and quality aspects of the system are considered and addressed \cite{aldhafer2022end,sommerville2011software}.

Requirements classification is integral to the success of software development \cite{aldhafer2022end,sommerville2011software,binkhonain2019review,cleland2007automated}. By improving the organization of requirements, it enables more effective management and prioritization \cite{sommerville2011software,riegel2015systematic,zave1997classification}.
Moreover, a well-structured classification enhances communication among stakeholders and ensures a unified, consistent understanding of the system’s needs \cite{sommerville2011software,zave1997classification,cleland2007automated}. 
This clarity is crucial for identifying potential conflicts or gaps early in the development lifecycle, thereby mitigating the risk of errors and contributing to the delivery of higher-quality software \cite{sommerville2011software,zave1997classification,cleland2007automated,hofmann2001requirements}. Additionally, requirements classification supports traceability, which is vital for ensuring compliance with regulatory and quality standards, thus reinforcing the integrity and accountability of the development process \cite{sommerville2011software,zave1997classification,cysneiros2001framework}.

A significant body of prior research has explored automated requirements classification using traditional machine learning models \cite{alhoshan2023zero,binkhonain2019review}. While these studies have yielded promising results, they suffer from two main limitations that stem from their general reliance on supervised learning \cite{alhoshan2023zero}. The first limitation is that supervised learning models require large volumes of annotated data for training. Acquiring annotated data is a challenging process that is often time-consuming, prone to errors, and requires expert involvement \cite{alhoshan2023zero,novielli2018benchmark,lin2018sentiment,zhang2023revisiting,lin2022opinion}. The second limitation is the lack of generalizability. Supervised models are often tailored to specific classification tasks and require retraining or fine-tuning with task-specific annotated data to achieve satisfactory performance. For example, a model trained to classify requirements according to one classification scheme cannot be directly applied to a different classification scheme; instead, a separate model trained on annotated data for the new classification is needed. This requirement for task-level customization further reinforces the dependency on annotated datasets, thereby increasing the resource demands of the development process \cite{alhoshan2023zero}.

The emergence of prompt-based large \acp{LLM} has transformed \ac{NLP} by enabling contextual understanding and task adaptability across a wide range of applications with minimal supervision. In contrast to traditional supervised approaches that require extensive annotated datasets and customization, prompt-based \acp{LLM} can be directed to perform diverse tasks by modifying only the input prompt. This capability substantially reduces or eliminates the dependence on task-specific annotated training data. Moreover, prompt-based \acp{LLM} exhibit strong generalizability across tasks and domains, as the core model remains fixed while only the prompt is adapted to the specific task at hand \cite{alammar2024hands,huyen2025ai}.

Despite their high potential, concerns persist regarding the cost, accessibility, privacy, and transparency of commercial or closed-source \acp{LLM} \cite{huyen2025ai,KIBRIYA2024109698}. Additionally, latency can be a bottleneck, especially when compared to lightweight, fine-tuned models that are optimized for specific tasks and deployed on low-resource devices. However, the growing availability of open-source models with a minimal performance gap compared to closed-source alternatives along with other cost-effective options, has alleviated many of these concerns and positioned prompt-based \acp{LLM} as a more practical and scalable solution for real-world applications \cite{huyen2025ai}.

While empirical evidence has highlighted the effectiveness of prompt-based \acp{LLM} across a range of domains, including healthcare \cite{10.1093/jamia/ocae145}, their performance has demonstrated notable limitations in others, such as the legal domain \cite{savelka2023unreasonable}. This variability in performance extends to the domain of software engineering. Although prompt-based \acp{LLM} have demonstrated strong results in relatively well-structured tasks, such as sentiment analysis \cite{zhang2023revisiting}, their utility diminishes when applied to more complex tasks, such as software vulnerability detection \cite{fu2023chatgpt}.

Building on the demonstrated success of prompt-based \acp{LLM} across various domains and software engineering tasks, the application of prompt-based \acp{LLM} for automated requirements classification provides a promising avenue to address some of its key challenges (i.e., the reliance on annotated data and the lack of generalizability). However, the effectiveness of prompt-based \acp{LLM} in this context remains uncertain due to their variable performance across domains and tasks. This uncertainty is further compounded by the unique linguistic and structural characteristics of requirements documents, which often include domain-specific terminology, ambiguous or imprecise phrasing, and need for fine-grained subclass distinctions. These factors highlight the importance of conducting a focused empirical investigation into the viability of prompt-based \acp{LLM} in requirements classification and identifying the prompting techniques that are most effective for requirements classification.

Therefore, this benchmarking study aims to investigate the effectiveness of prompt-based \acp{LLM} for the automatic classification of software engineering requirements. We evaluate and statistically compare the performance of five prompt-based \acp{LLM} across three requirements classification tasks using four promoting techniques. We broaden the scope of our analysis by benchmarking the performance of prompt-based \acp{LLM} against a state-of-the-art fine-tuned transformer model.

The main contributions of this study can be summarized as follows: \begin{itemize}
 \item A comparative evaluation of five prompt-based \acp{LLM} using zero-shot prompting across three requirements classification tasks.
\item A comparative evaluation of the five prompt-based  \acp{LLM} using few-shot prompting across the three tasks.
\item An investigation of the impact of augmenting both zero-shot and few-shot prompts with persona and \ac{CoT} techniques on classification performance.
\item A benchmarking of the best-performing prompt-based \acp{LLM} and prompts configurations against a state-of-the-art fine-tuned transformer model across all three tasks.
\item Empirical evidence that prompt-based \acp{LLM} can reduce reliance on large annotated datasets while achieving competitive performance on the three classification tasks. 

\end{itemize}

The remainder of this paper is organized as follows: Section \ref{Related work} summarizes previous, related work. Section \ref{Study setup} outlines the study setup. Section \ref{Results} presents the results, and Section \ref{Discussion} discusses them. Section \ref{Threats to validity} addresses potential threats to the study’s validity, and Section \ref{Conclusion} concludes the study.

\section{Related work} \label{Related work}
Requirements classification has been one of the interests of the software engineering community. Binkhonain and Zhao \cite{binkhonain2019review} conducted a systematic review on requirements classification that is focused on the identification of \acp{NFR}, and we encourage readers to consult their work for a comprehensive summary of current approaches. Below, we highlight the studies most closely related to our research.

Research in this area has largely focused on distinguishing between \acp{FR} and \acp{NFR} and further classifying various types of \acp{NFR}. Cleland-Huang et al. \cite{cleland2007automated} introduced the PROMISE \ac{NFR} dataset, which has since become a widely used benchmark in the research community and also serves as one of the benchmarks in our work. In their study, they proposed a supervised approach for identifying different \ac{NFR} classes using a set of indicator terms. These terms were extracted from manually annotated requirements and then applied to classify unseen requirements. While the approach achieved a recall of up to 0.80, it suffered from low precision, reaching only 0.21.

To address the challenge of dataset annotation, Casamayor et al. \cite{casamayor2010identification} developed a semi-supervised Naive Bayes approach built on an iterative process similar to active learning, where the user provides feedback to the classifier. Using the PROMISE \ac{NFR} dataset as training data, their approach, after multiple iterations, achieved maximum precision above 0.80 and maximum recall above 0.70 for most classes, except for underrepresented ones.

Kurtanović and Maalej \cite{8049171} developed an \ac{SVM} model that combined metadata, lexical, and syntactical features to classify requirements as \ac{FR} or \ac{NFR} and further classify \ac{NFR} types. Their approach achieved up to 92\% precision and recall for classfying \ac{FR} and \ac{NFR}, and for specific \ac{NFR} classes, the highest precision (92\%) and recall (90\%) were observed for security and performance requirements.

Dalpiaz et al. \cite{dalpiaz2019requirements} proposed an interpretable machine learning approach to address the challenge of classifying large datasets of requirements. They introduced 17 general linguistic features, including dependency types, to build transparent models. Using modern introspection tools, they engineered these features to expose the classifier’s inner workings. While their models fit the training data slightly less well than dense feature models, they showed better generalization on validation data and improved interpretability for human analysts.

Navarro-Almanza et al. \cite{navarro2017towards} proposed a deep learning–based approach to reduce the substantial human effort typically needed for manual feature engineering. Rather than relying on traditional \ac{NLP} or information retrieval techniques, they applied a \ac{CNN}, leveraging its success in other \ac{NLP} tasks. On the PROMISE dataset, their model achieved a precision of 0.80 and a recall of 0.79.

Similarly, Aldhafer et al. \cite{aldhafer2022end} proposed an end-to-end deep learning approach using Bidirectional Gated Recurrent Units (BiGRU) to classify requirements directly from raw text, using either word or character sequences as input. Their study found that word-based models generally outperformed character-based ones and surpassed the performance of many traditional and deep learning baselines.

Hey et al. \cite{9218141} fine-tuned the BERT model on the PROMISE dataset, achieving 0.92 precision and 0.95 recall for the classification of \acp{FR} and \acp{NFR}. For \ac{NFR} classification, the fine-tuned model reached precision of up to 0.94 and recall of up to 0.90. The model was also successfully applied to classify \acp{FR} into subclasses, achieving precision of up to 0.88 and recall of up to 0.95.

Other several studies have specifically addressed security requirements classification. Knauss et al. \cite{knauss2011supporting} employed a Bayesian classifier to identify security-related requirements across three industrial datasets, achieving precision above 0.80 and recall above 0.90. Riaz et al. \cite{6912260} developed a semi-automated K-nearest neighbors (KNN) approach to extract security-relevant sentences from requirements documents, using a dataset of 10,963 sentences from six healthcare domain documents; they reported a precision of 0.82 and recall of 0.79. To address the lack of domain-specific datasets, Munaiah et al. \cite{munaiah2017domain} designed a domain-independent one-class \ac{SVM} classifier to identify general software security weakness descriptions, trained on the Common Weakness Enumeration (CWE) database \cite{christey2013common}. The classifier achieved an average precision of 0.67 and recall of 0.70. More recently, Varenov et al. \cite{varenov2021security} compared several fine-tuned transformer models (i.e., BERT, XLNet, and DistilBERT) to classify security requirements into sub-classes such as confidentiality, integrity, availability, accountability, operational, access control, and others. DistilBERT achieved the best performance, with a precision of 0.80 and recall of 0.82.

A number of studies have addressed the classification of \ac{FR} and \ac{NFR} as well as security requirements classification, rather than focusing solely on one. 
For instance, Dekhtyar and Fong \cite{dekhtyar2017re} used Word2Vec embeddings combined with \acp{CNN} to classify requirements. Compared to baseline Naïve Bayes classifiers on the SecReq and \ac{NFR} datasets, their approach achieved an F1-score improvement of over 7\% on the SecReq dataset and approximately 6\% on the \ac{NFR} dataset. To address the scarcity of annotated datasets, Alhoshan et al. \cite{alhoshan2023zero} explored the potential of embedding-based zero-shot learning for requirements classification. They evaluated multiple models and labels across the following three classification tasks: distinguishing between \acp{FR} and \acp{NFR}, identifying specific \acp{NFR} classes, and distinguishing between security and non-security requirements. The best F1 scores achieved for these tasks are lower than the results reported by the supervised approaches of Kurtanović and Maalej \cite{8049171} and Hey et al. \cite{9218141}. 

While this study addresses the same three classification tasks and shares Alhoshan et al.’s objective of minimizing reliance on annotated data typically required for supervised learning, it adopts a distinct approach. Specifically, this study leverages prompt-based \acp{LLM} and prompting techniques in contrast to the embedding-based approach employed by Alhoshan et al. \cite{alhoshan2023zero}.

\section{Study setup} \label{Study setup}
This section outlines the benchmarking study setup, including its goals and research questions (RQs), datasets, tasks, selected \acp{LLM}, prompts, performance measures, statistical analysis, and implementation details.

We frame our study as a benchmarking study, in alignment with the ACM SIGSOFT Empirical Standards for Software Engineering \cite{ralph2020empirical}. Benchmarking is widely used in computer science to compare the performance of algorithms, tools, and platforms in a repeatable and objective manner. It involves standardized measurements and datasets to enable fair, reproducible, and comparable assessments across different techniques, and is commonly adopted in areas, such as database systems, information retrieval, and cloud computing \cite{10.1145/3463274.3463361,ralph2020empirical}. 

A benchmarking study is particularly well-suited to this study’s setting, where the goal is to compare the performance of prompt-based \acp{LLM}, various prompting techniques, and a state-of-the-art fine-tuned transformer-based model using fixed datasets and a performance measure across multiple requirements classification tasks.

\subsection{Goals and Research Questions (RQs)}
The goal of this study is formally defined using the Goal-Question-Metric (GQM) template \cite{caldiera1994goal}, as follows: Analyze the effectiveness of prompt-based \acp{LLM}
for the purpose of classifying software requirements
with respect to their classification performance
from the viewpoint of software engineers and researchers
in the context of applying four distinct prompting techniques on selected datasets across three classification tasks, and benchmarking their classification performance against that of a state-of-the-art fine-tuned transformer-based model.

To achieve this goal, we derived four research questions (RQs), which are presented below along with the motivation behind each RQ.

\textbf{Motivation:} As mentioned previously, the advancement of prompt-based \acp{LLM} across a range of domains and various software engineering tasks has highlighted their potential for requirements classification \cite{zhang2023revisiting,10.1093/jamia/ocae145}. A key advantage of these models is their ability to operate in a zero-shot learning, eliminating the need for annotated training data \cite{alammar2024hands}. Zero-shot prompting is particularly appealing to practitioners, as it enables rapid utilization without requiring domain-specific tuning. However, the effectiveness of such models in the context of requirements classification remains largely unexplored. Therefore, this study aims to gain a deeper understanding of how these models perform under a zero-shot configuration for requirements classification. This leads to the study’s first RQ being the following: 
 \textbf{RQ1: How effective are prompt-based \acp{LLM} when applied to software requirements classification in a zero-shot configuration?}

\textbf{Motivation:} Few-shot learning enables in-context learning by incorporating a small number of labeled examples directly into the prompt, eliminating the need for formal training, fine-tuning, or large annotated datasets \cite{alammar2024hands}. While it is reasonable to expect that such task-specific examples can enhance model performance, the extent and nature of this impact remain unclear. This calls for an investigation focused on requirements classification, where the complexity and diversity of classes may influence the effectiveness of provided examples. Therefore, the second research question of this study is as follows:
\textbf{RQ2: How effective are prompt-based \acp{LLM} when applied to software requirements classification in a few-shot configuration?}

\textbf{Motivation:} Prompt framing strategies play a critical role in shaping how a prompt-based \ac{LLM} interprets and generates responses. Techniques such as persona-based prompting and \ac{CoT} reasoning are designed to influence the model’s behavior by assigning it a specific role or encouraging structured, step-by-step reasoning \cite{alammar2024hands}. While these techniques have demonstrated effectiveness in other domains, their impact on requirements classification remains unexplored. Given the complexity and domain-specific nature of software requirements, it is important to assess whether such strategies can improve model performance in this context. Therefore, the third RQ of this study is as follows:
\textbf{RQ3: How do prompt framing strategies, specifically personas and \ac{CoT}, affect the classification performance of prompt-based \acp{LLM}?}

\textbf{Motivation:} 
While prompt-based \acp{LLM} provides the advantage of minimal setup and flexibility across tasks \cite{alammar2024hands}, fine-tuned transformer-based models remain the dominant approach for requirements classification due to their strong performance on benchmark datasets \cite{alhoshan2023zero,9218141}. These models, however, rely on large volumes of annotated data and require retraining or fine-tuning for each specific task \cite{alhoshan2023zero}, which may limit their practicality in real-world settings. In contrast, prompt-based \acp{LLM} approaches eliminate the need for fine-tuning, but their effectiveness relative to established supervised baselines remains unclear. To assess the viability of prompt-based \acp{LLM} as a competitive alternative, it is important to compare their performance against state-of-the-art fine-tuned transformer-based model. Therefore, the fourth RQ of this study is as follows: \textbf{RQ4: How does the performance of prompt-based \acp{LLM} compare to the state-of-the-art fine-tuned transformer-based model for software requirements classification?
}

\subsection{Datasets}
To address the RQs, we utilized two publicly available datasets that are commonly used for benchmarking requirements classification in software engineering \cite{alhoshan2023zero,dekhtyar2017re}.

The first dataset is the PROMISE \ac{NFR} dataset, which contains a total of 625 software requirements, divided into 255 \acp{FR} and 370 \acp{NFR}. The \acp{NFR} are further classified into 11 sub-classes: availability (21), legal (13), look and feel (38), maintainability (17), operational (62), performance (54), scalability (21), security (66), usability (67), fault tolerance (10), and portability (1) \cite{jane_cleland_huang_2007_268542,cleland2007automated}. In this study, we excluded the portability class due to it having only a single instance.

The second dataset is SecReq, which consists of 510 software requirements, with 187 classified as security-related requirements and the remaining 323 classified as non-security requirements \cite{knauss_2021_4530183,knauss2011supporting}.

\subsection{Tasks}
As the primary objective of this study is to examine the potential of prompt-based \acp{LLM} in automatically classifying software requirements, the core task we evaluate involves the goal of classifying natural language requirements into their respective class types. Specifically, our empirical investigation focuses on the following three classification tasks that are widely studied within software engineering research \cite{alhoshan2023zero,dekhtyar2017re}:

\begin{itemize}
 \item \textbf{Binary classification of \acp{FR} and \acp{NFR} (FR-NFR):} This task aims to classify a given requirement as either a \ac{FR} or a \ac{NFR}. For ease of reference, this task will be referred to as FR-\ac{NFR} in the remainder of the paper. The PROMISE dataset is used for this task.

 \item \textbf{Multi-class classification of \acp{NFR} (MC-NFR):} This task aims to classify a given \ac{NFR} into one of the 10 retained \ac{NFR} classes (i.e., the portability class was excluded as it only had one instance). It is a single-label classification task, meaning each \ac{NFR} is assigned to only one class. This task will be referred to as MC-\ac{NFR} in the remainder of the paper, and the PROMISE dataset is used for this task.

 \item \textbf{Binary classification of security and non-security requirements (Sec-NonSec):} This task aims to classify a given requirement as either security-related (Sec) or non-security-related (NonSec). This task will be referred to as Sec-NonSec in the remainder of the paper. The SeqReq dataset is used for this task.
\end{itemize}

\subsection{Selected \acp{LLM} }
To conduct our study, we selected the following five models:
\begin{itemize} 

\item Claude: \footnote{\url{https://pypi.org/project/anthropic/}}: Claude is a family of \acp{LLM} developed by Anthropic. For this study, we used the Claude 3 Haiku variant, which is designed for lightweight, high-speed inference. Claude 3 Haiku is a closed-source model that is accessible through Anthropic’s paid application programming interface (API).

\item DeepSeek \footnote{\url{https://api.deepseek.com}}: DeepSeek is a family of \acp{LLM} developed by the DeepSeek company. We used DeepSeek-V3, a state-of-the-art Mixture-of-Experts (MoE) model that is open weight and accessible via a paid API.

\item Gemini \footnote{\url{https://ai.google.dev/gemini-api/docs/models}}: Gemini is a family of models developed by Google. In this study, we used Gemini 2.0 Flash, a variant optimized for both speed and performance. 
Gemini 2.0 Flash is closed-source model that is accessible through a free-tier API with usage limits and supports higher-volume usage via a paid, pay-as-you-go API. We note that the experiments in this study remained within the free-tier limits and did not incur any charges.

\item GPT-4 \footnote{\url{https://platform.openai.com/docs/models/gpt-4}}: GPT is a family of \acp{LLM} developed by OpenAI. For this study, we used GPT-4 Turbo, the most advanced variant available at the time of the experiment. GPT-4 Turbo is a closed-source model, accessible through a paid API provided by OpenAI.

\item Llama \footnote{\url{https://huggingface.co/meta-llama/Llama-3.2-3B-Instruct}}: Llama  is a family of \acp{LLM} developed by Meta. In this study, we used the Llama -3.2 3B Instruct variant, a compact model designed for fast and accurate performance. It is an open-weight model and is accessible through a free API.

\end{itemize}

All models were accessed through their publicly available APIs. We used the default parameters provided by each model’s API, with two exceptions. First, we standardized the context length to 8,192 tokens across all models to avoid any advantage due to differences in maximum context window sizes. Second, we set the temperature to zero to reduce variability in model outputs.

\subsection{Prompts}
Prompts play a crucial role in the utilization of prompt-based \acp{LLM}, as their structure and wording can significantly influence model performance \cite{huyen2022designing}. Therefore, the first author crafted the initial prompts based on the prompt engineering guidelines \cite{huyen2022designing,alammar2024hands}. These prompts were subsequently reviewed by the second author, and no modifications were deemed necessary after the review.

To evaluate the performance of the selected \acp{LLM} on the requirements classification tasks, we utilized the following four prompting techniques: zero-shot, few-shot, persona, and \ac{CoT}. We describe each prompt technique below, and Table \ref{tab:prompt_categories} provides a representative example for each. The full set of prompts used for each task is available in our online appendix \footnote{\url{https://osf.io/2zbcu/?view_only=09681b1cf4b845c9b771d9107246f15f}}

\begin{itemize}
 \item Zero-shot: In zero-shot prompting, the model receives only a natural language description of the task, without any examples or demonstrations \cite{alammar2024hands}.

\item Few-shot prompts include the task description along with a small number of examples to guide the model’s output \cite{alammar2024hands}. Although various techniques exist for selecting these examples \cite{liu-etal-2022-makes}, we opted for random selection in this study because our goal was to assess whether simply providing examples improves performance, rather than optimizing example selection using techniques, such as embedding similarity. 
Additionally, we did not seek to have the selected examples be representative of all classes in order to maintain a fair comparison, as providing one example per class for the MC-NFR task would require including ten examples. Including such a large number of examples could overwhelm the prompt, which would shift the focus away from our primary goal of evaluating whether the presence of examples, regardless of the quality of their content, improves performance.

 \item Persona: This technique augments both zero-shot and few-shot prompts with role-specific instructions to assess whether such contextual framing influences performance \cite{huyen2022designing}. 

\item \Acf{CoT}: This technique augments both zero-shot and few-shot prompts with instructions for the model to provide intermediate reasoning steps. The goal is to assess whether structured reasoning improves the model's decision-making and performance on the task \cite{huyen2022designing}.
\end{itemize}

\begin{table}[htbp]
\footnotesize

\caption{Summary of the prompting techniques used in the study, along with an example for each technique}
\label{tab:prompt_categories}
\begin{tabular}{@{}p{4cm} p{11cm}@{}}
\toprule
\textbf{Prompting technique} & \textbf{Example} \\ \midrule

\multirow{3}{4cm}{Zero-shot} 
& You will be provided with a requirement statement, and your task is to classify it as either Functional or Non-functional. Return only the class label, with no additional text. \\
& Requirement: \{req\_text\} \\
& Class → \\ \midrule

\multirow{6}{4cm}{Few-shot} 
& You will be provided with a requirement statement, and your task is to classify it as either Functional or Non-functional. Return only the class label, with no additional text. \\
& Here are some examples: \\
& Requirement 1: The product will be able to delete conference rooms. \\
& Class → Functional \\

& Now classify: \{req\_text\} \\
&Class → \\ \midrule

\multirow{3}{4cm}{Persona} 
& Act as an experienced Requirements Analyst and classify the requirement as either Functional or Non-functional. Return only the class label, with no additional text. \\
& Requirement: \{req\_text\} \\
&Class → \\ \midrule

\multirow{5}{4cm}{\Acf{CoT}} 
&You will be provided with a requirement statement, and your task is to classify it as either Functional or Non-functional. Think step by step before assigning a label.\\
& Requirement: \{req\_text\}
\\
&Class → \\ 
\bottomrule
\end{tabular}
\end{table}

\subsection{Performance measure}
To evaluate the performance of the examined models and prompts, we calculated a range of measures, such as precision, recall, F1-score, weighted F1-score, and micro F1-score. However, due to space limitations, we report only the macro-F1 score in this paper; the complete set of evaluation results is available in our online appendix.

Our decision to focus primarily on the macro-F1 score is motivated by the characteristics of the datasets used in this study. Given their high class imbalance (i.e., the number of instances varies significantly across classes), it is essential to select performance measures meticulously, as some measures may produce misleading results.

For instance, accuracy is often biased toward the majority classes \cite{manning2008introduction}. Moreover, although the weighted F1-score accounts for class distribution, it can still be dominated by performance on the majority class, which results in poor performance on minority classes \cite{10.1162/tacl_a_00675}. Similarly, the micro-F1 score may lead to misleading assessments at the class level, as it assigns equal weight to each individual prediction rather than to each class. This often yields poor visibility of minority class performance, since the measure tends to favor majority classes \cite{manning2008introduction}.

In contrast, the macro-F1 score treats all classes equally by computing the unweighted average of the F1-scores for each class; as a result, it is considered a more appropriate and reliable measure for evaluating model performance on imbalanced datasets \cite{manning2008introduction}. The utilization of macro-F1 score is consistent with prior work  \cite{novielli2020can, zhang2023revisiting}. Below, we present the definitions and formulas for the macro-F1 score, the F1-score, and their component metrics (i.e., precision and recall). 

\begin{itemize}

\item \textbf{Macro-F1}: The macro-F1 score is the average of the independent F1 scores across all classes. It provides an overall measure of the model's performance, giving equal weight to each class regardless of its frequency. It is calculated as follows:
\begin{equation*}
\text{Macro-F1} = \frac{1}{N} \sum_{i=1}^{N} F1_i
\end{equation*}

Where:
\begin{itemize}
 \item $N$ is the total number of classes.
 \item $F1_i$ is the F1 score for class $i$.
\end{itemize}

\item \textbf{F1-score}: The F1 score is the harmonic mean of precision and recall, balancing both metrics to account for false positives and false negatives. It is calculated as follows:
\begin{equation*}
F1 = 2 \times \frac{\text{precision} \times \text{recall}}{\text{precision} + \text{recall}}
\end{equation*}

Where:
\begin{itemize}
 \item \textbf{Precision}: measures the accuracy of the model’s positive predictions, and it is calculated as follows:
 \begin{equation*}
 \text{Precision} = \frac{TP}{TP + FP}
 \end{equation*}
 
 \item \textbf{Recall}: measures the completeness of the model’s positive predictions, and it is calculated as follows:
 \begin{equation*}
 \text{Recall} = \frac{TP}{TP + FN}
 \end{equation*}

 \item $TP$ refers to true positives cases where the model correctly predicts the positive class.
 \item $FP$ refers to false positives cases where the model incorrectly predicts the positive class.
 \item $TN$ refers to true negatives cases where the model correctly predicts the negative class.
 \item $FN$ refers to false negatives cases where the model incorrectly predicts the negative class.
\end{itemize}

\end{itemize}

\subsection{Statistical analysis}
Merely comparing performance measure values is not sufficient, as observed differences may arise due to random variability \cite{witten2005practical}. To assess the significance of these differences, a statistical analysis is necessary. Therefore, we evaluate the performance of the classification models using the non-parametric Scott-Knott Effect Size Difference (ESD) test. We opted to use this test due to its ability to produce distinct, non-overlapping groups, as opposed to other tests, such as Nemenyi’s test that may result in overlapping groupings. Moreover, not only the Scott-Knott ESD test detects statistical differences, but it also quantifies the magnitude of median differences and only considers meaningful differences \cite{tantithamthavorn2016empirical, tantithamthavorn2018impact}. Additionally, as a non-parametric test, the test is robust against outliers and holds no assumptions regarding homogeneity, normality, or sample size \cite{puth2015effective, abdi2007kendall}. Lastly, the test has been successfully applied in similar contexts in previous studies \cite{tantithamthavorn2016empirical, tantithamthavorn2018impact}.

The Scott-Knott ESD test has two main steps. The first step focuses on identifying a partition that maximizes the differences between group medians. To achieve this, the test first sorts the median values and then applies the non-parametric Kruskal-Wallis Chi-square test to find a partition that maximizes median differences. The second step iteratively splits or merges groups based on the statistical significance and magnitude of the median differences using Cliff’s $|\delta|$ effect size. If the difference between the medians of two groups is considered statistically significant and non-negligible, they are separated; otherwise, they are merged \cite{tantithamthavorn2016empirical, tantithamthavorn2018impact}.
\subsection{Implementation} 
To conduct the empirical assessment, we first partitioned each dataset into training, validation, and test sets using an 8:1:1 ratio with stratified splitting. For RQ1, RQ2, and RQ3, we evaluated all models using their associated prompts on the test split associated with each task. We restricted our evaluation to the test sets to ensure fair comparisons with RQ4, which is consistent with the approach adopted in prior related studies \cite{zhang2023revisiting,Tawosi}. 

To address RQ4, we fine-tuned BERT, a transformer-based model recognized as state-of-the-art by \cite{alhoshan2023zero,9218141} for requirement classification. The model was fine-tuned using a learning rate of $2 \times 10^{-5}$, for 5 epochs, with a batch size of 32 and a maximum sequence length of 256 tokens. The model achieving the highest macro-F1 score on the validation set was then evaluated on the test set of each task. We compared the performance of the fine-tuned model against the best-performing model–prompt configurations from the earlier RQs for each requirements classification task.

\section{Results} \label{Results}
This section summarizes the results of the study. 

\subsection{RQ1: How effective are prompt-based \acp{LLM} when applied to software requirements classification in a zero-shot configuration?}

The results of zero-shot prompting for each model across the three requirement classification tasks are presented in Table \ref{tab:RQ1}, where the highest macro-F1 score is bolded and ``\#ZS\'' refers to zero-shot prompt. 

For the FR-\ac{NFR} task, Claude achieved the highest macro-F1 score (0.73), followed by Llama (0.67), Gemini (0.65), DeepSeek (0.57), and GPT4 (0.53).

In the MC-\ac{NFR} task, DeepSeek achieved the highest macro-F1 score at 0.88, followed by Gemini with 0.83 score. GPT4 obtained a score of 0.82, and Claude has a score of .75. Llama preformed the worst with a score of 0.68.

For the Sec-NonSec task, GPT4 led with a macro-F1 score of 0.73, followed by Gemini (.68), Llama (0.58), Claude (.57), and DeepSeek (0.48).

\begin{table}[htbp]
\centering
\caption{Zero-shot macro-F1 scores of \acp{LLM} across software requirement classification tasks}
\label{tab:RQ1}
\begin{tabular}{@{}llll@{}}
\toprule
Model-prompt & FR-\ac{NFR}& MC-\ac{NFR}& Sec-NonSec \\ \midrule
Claude\_ZS & \textbf{0.73} & 0.75 & 0.57 \\
DeepSeek\_ZS & 0.57 & \textbf{0.88} & 0.48 \\
Gemini\_ZS & 0.65 & 0.83 & 0.68 \\
GPT4\_ZS & 0.53 & 0.82 & \textbf{0.73} \\
Llama\_ZS & 0.67 & 0.68 & 0.58 \\ \bottomrule
\end{tabular}
\end{table}

\begin{figure} [hbtp]
 \center
 \includegraphics[width=.65
 \textwidth]{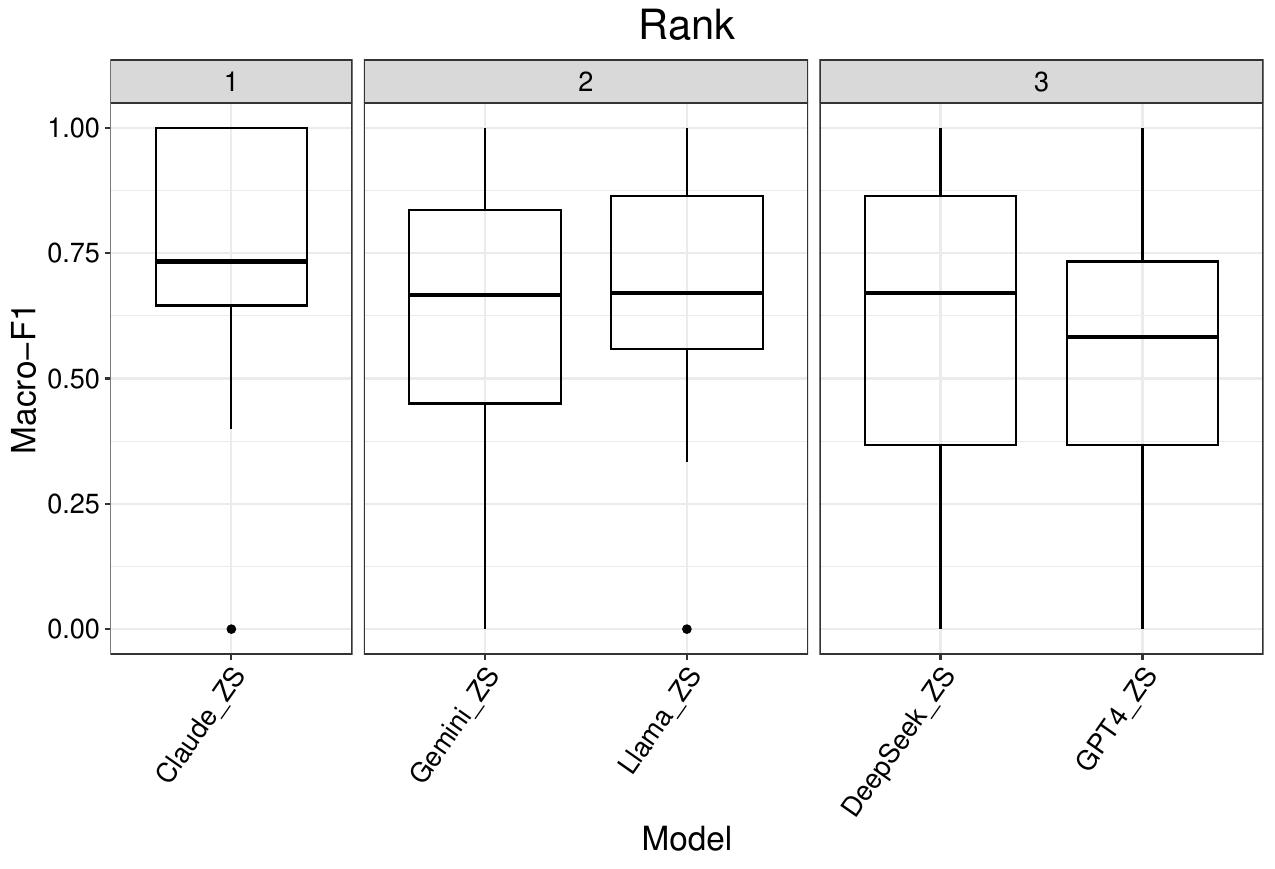}
\caption{Scott-Knott ESD ranking of zero-shot prompting for each \ac{LLM} on the FR-\ac{NFR} task based on macro-F1 score}
 \label{fig:RQ1.1}
\end{figure}

The results of the statistical analysis for the FR-\ac{NFR} task are presented in Figure \ref{fig:RQ1.1}. The test identified Claude as the top performer, followed by Gemini and Llama in the second rank, while both DeepSeek and GPT were ranked third.

\begin{figure} [hbtp]
 \center
 \includegraphics[width=.65
 \textwidth]{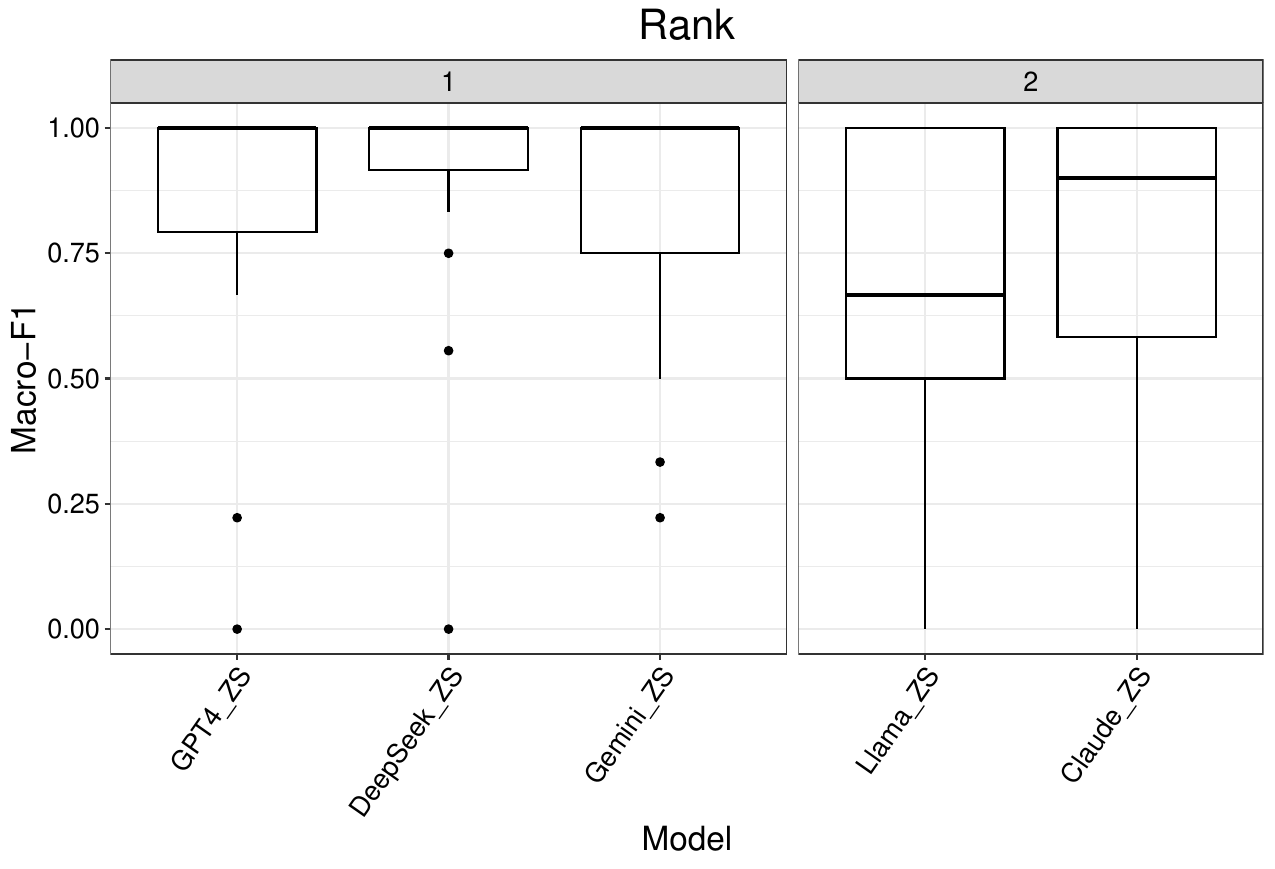}
\caption{Scott-Knott ESD ranking of zero-shot prompting for each \ac{LLM} on the MC-\ac{NFR} task based on macro-F1 score}
 \label{fig:RQ1.2}
\end{figure}

For the MC-\ac{NFR} task, as presented in Figure \ref{fig:RQ1.2}, the analysis revealed two performance ranks: GPT4, DeepSeek, and Gemini were ranked as the top-performing models, while Llama and Claude formed the second rank.

For the Sec-NonSec task, the test ranked GPT4 as the best performer, followed by Gemini in second place, both Claude and Llama are ranked third rank, and DeepSeek was ranked fourth.

\begin{figure} [hbtp]
 \center
 \includegraphics[width=.65
 \textwidth]{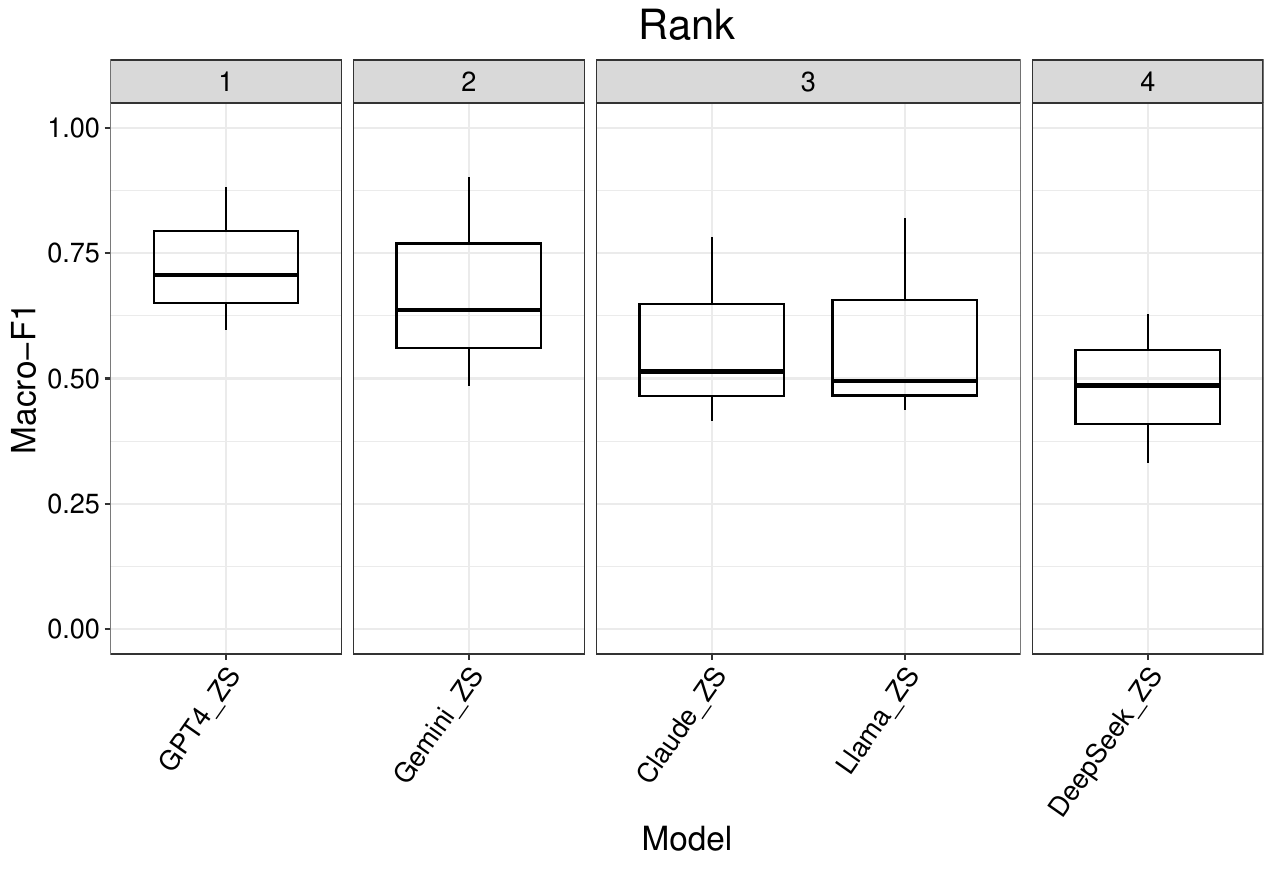}
\caption{Scott-Knott ESD ranking of zero-Shot prompting for each \ac{LLM} on the Sec-NonSec task based on macro-F1 score}
 \label{fig:RQ1.3}
\end{figure}

\subsection{RQ2: How effective are prompt-based \acp{LLM} when applied to software requirements classification in a few-shot configuration?}

The resulting macro-F1 scores of applying the few-shot prompts using each model across the three software requirement classification tasks are presented on 
Table \ref{tab:RQ2}. The highest value for each task is bolded, and the ``\#FS\'' denotes the number of few-shot examples provided, where ``\#'' represents the specific number of examples used in the prompt. Moreover, an upward arrow ($\uparrow$) indicates an increase in the obtained macro-F1 score compared to the zero-shot prompt, a downward arrow ($\downarrow$) indicates a decreased value, and an equals sign ($=$) indicates no change.

In the FR-\ac{NFR} task, Gemini\_5FS achieved the highest macro-F1 score of 0.92, followed by DeepSeek\_5FS, Gemini\_3FS, and GPT4\_5FS, with scores of 0.88. On the other hand, Llama few-shot configurations consistently demonstrated weak performance, with Llama\_1FS scoring the lowest at 0.40, followed by Llama\_3FS (0.53) and Llama\_5FS (0.61). Although the performance of the majority of models increased with few-shot prompts, Llama struggled when provided with examples. Claude's performance has also degraded when provided with one example, but it improved with both three-shot and five-shot prompts.

Similarly, Gemini\_5FS performed the best on the MC-\ac{NFR} task, achieving the highest macro-F1 score of 0.9, followed closely by Gemini\_1FS with a score of 0.89. This was followed by Claude\_3FS, Claude\_5FS, DeepSeek\_1FS, DeepSeek\_3FS, and DeepSeek\_5FS, all with a macro-F1 score of 0.85. Llama few-shot prompts configuration were among the weakest performers, with Llama\_1FS achieving the lowest score of 0.75, followed by both Llama\_3FS and Claude\_1FS with scores of 0.78, and Llama\_5FS with a score of 0.80. Although the highest obtained macro-F1 score was achieved through five-shot learning, providing examples for this specific task had no influence on some models, such as GPT4.

In the Sec-NonSec task, GPT4\_5FS achieved the highest macro-F1 score of 0.87, followed closely by Gemini\_5FS with 0.85 and Gemini\_1FS with 0.83. Llama models demonstrated the weakest performance overall, with Llama\_1FS scoring 0.65, Llama\_5FS at 0.48, and Llama\_3FS performing the worst with a score of 0.43. The inclusion of examples in the prompts for this task had a positive effect on the performance of all models, except for Llama, whose performance decreased when using three-shot and five-shot prompts.

\begin{table}[hbtp]
\caption{Few-shot macro-F1 scores of \acp{LLM} across software requirement classification tasks}
\label{tab:RQ2}
\begin{tabular}{@{}llll@{}}
\toprule
Model-prompt & FR-\ac{NFR}& MC-\ac{NFR}& Sec-NonSec \\ \midrule
Claude\_1FS & 0.66\textsuperscript{$\downarrow$} & 0.78\textsuperscript{$\uparrow$} & 0.72\textsuperscript{$\uparrow$} \\
Claude\_3FS & 0.87\textsuperscript{$\uparrow$} & 0.85\textsuperscript{$\uparrow$} & 0.66\textsuperscript{$\uparrow$} \\
Claude\_5FS & 0.86\textsuperscript{$\uparrow$} & 0.85\textsuperscript{$\uparrow$} & 0.77\textsuperscript{$\uparrow$} \\
DeepSeek\_1FS & 0.75\textsuperscript{$\uparrow$} & 0.85\textsuperscript{$\downarrow$} & 0.79\textsuperscript{$\uparrow$} \\
DeepSeek\_3FS& 0.87\textsuperscript{$\uparrow$} & 0.85\textsuperscript{$\downarrow$} & 0.76\textsuperscript{$\uparrow$} \\
DeepSeek\_5FS& 0.88\textsuperscript{$\uparrow$} & 0.85\textsuperscript{$\downarrow$} & 0.79\textsuperscript{$\uparrow$} \\
Gemini\_1FS& 0.82\textsuperscript{$\uparrow$} & 0.89\textsuperscript{$\uparrow$} & 0.83\textsuperscript{$\uparrow$} \\
Gemini\_3FS& 0.88\textsuperscript{$\uparrow$} & 0.83\textsuperscript{$=$} & 0.82\textsuperscript{$\uparrow$} \\
Gemini\_5FS& \textbf{0.92}\textsuperscript{$\uparrow$} & \textbf{0.9}\textsuperscript{$\uparrow$} & 0.85\textsuperscript{$\uparrow$} \\
GPT4\_1FS& 0.64\textsuperscript{$\uparrow$} & 0.82\textsuperscript{$=$} & 0.81\textsuperscript{$\uparrow$} \\
GPT4\_3FS & 0.81\textsuperscript{$\uparrow$} & 0.82\textsuperscript{$=$} & 0.81\textsuperscript{$\uparrow$} \\
GPT4\_5FS& 0.88\textsuperscript{$\uparrow$} & 0.82\textsuperscript{$=$} & \textbf{0.87}\textsuperscript{$\uparrow$} \\
Llama\_1FS& 0.4\textsuperscript{$\downarrow$} & 0.75\textsuperscript{$\uparrow$} & 0.65\textsuperscript{$\uparrow$} \\
Llama\_3FS& 0.53\textsuperscript{$\downarrow$} & 0.76\textsuperscript{$\uparrow$} & 0.43\textsuperscript{$\downarrow$} \\
Llama\_5FS& 0.61\textsuperscript{$\downarrow$} & 0.8\textsuperscript{$\uparrow$} & 0.48\textsuperscript{$\downarrow$} \\ \bottomrule
\end{tabular}
\end{table}

\begin{figure} [hbtp]
 \center
 \includegraphics[width=.98
 \textwidth]{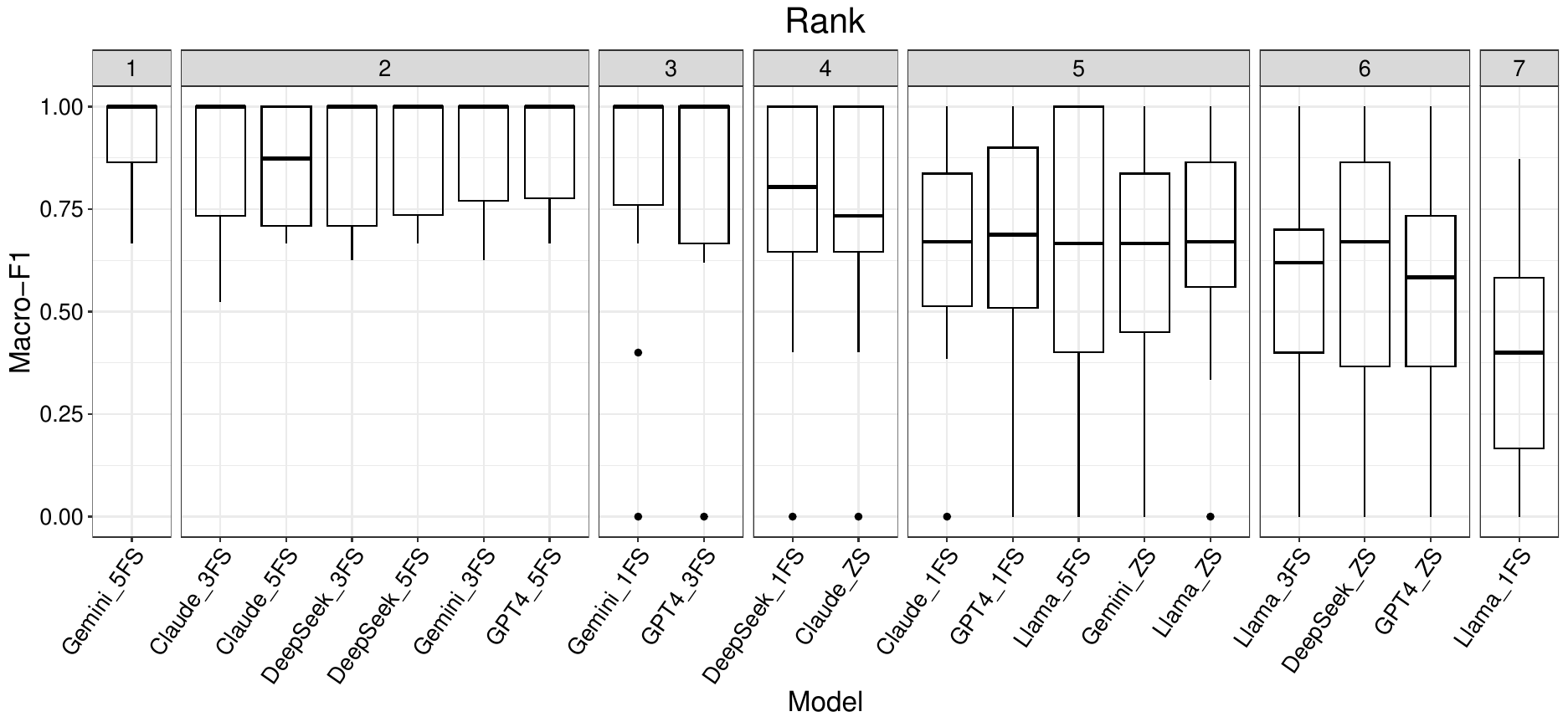}
\caption{Scott-Knott ESD ranking of zero-shot and few-shot promptings for each \ac{LLM} on the FR-\ac{NFR} task based on macro-F1 score}
 \label{fig:RQ2.1}
\end{figure}

The results of the statistical comparison of model-prompt configuration for the FR-\ac{NFR} task using both zero-shot and few-shot promptings are presented in Figure \ref{fig:RQ2.1}. The rankings indicate that the best-performing model-prompt configuration is Gemini\_5FS. Moreover, Claude\_3FS, Claude\_5FS, DeepSeek\_3FS, DeepSeek\_5FS,
Gemini\_3FS, and GPT4\_5FS configurations were ranked second.

\begin{figure} [hbtp]
 \center
 \includegraphics[width=.98
 \textwidth]{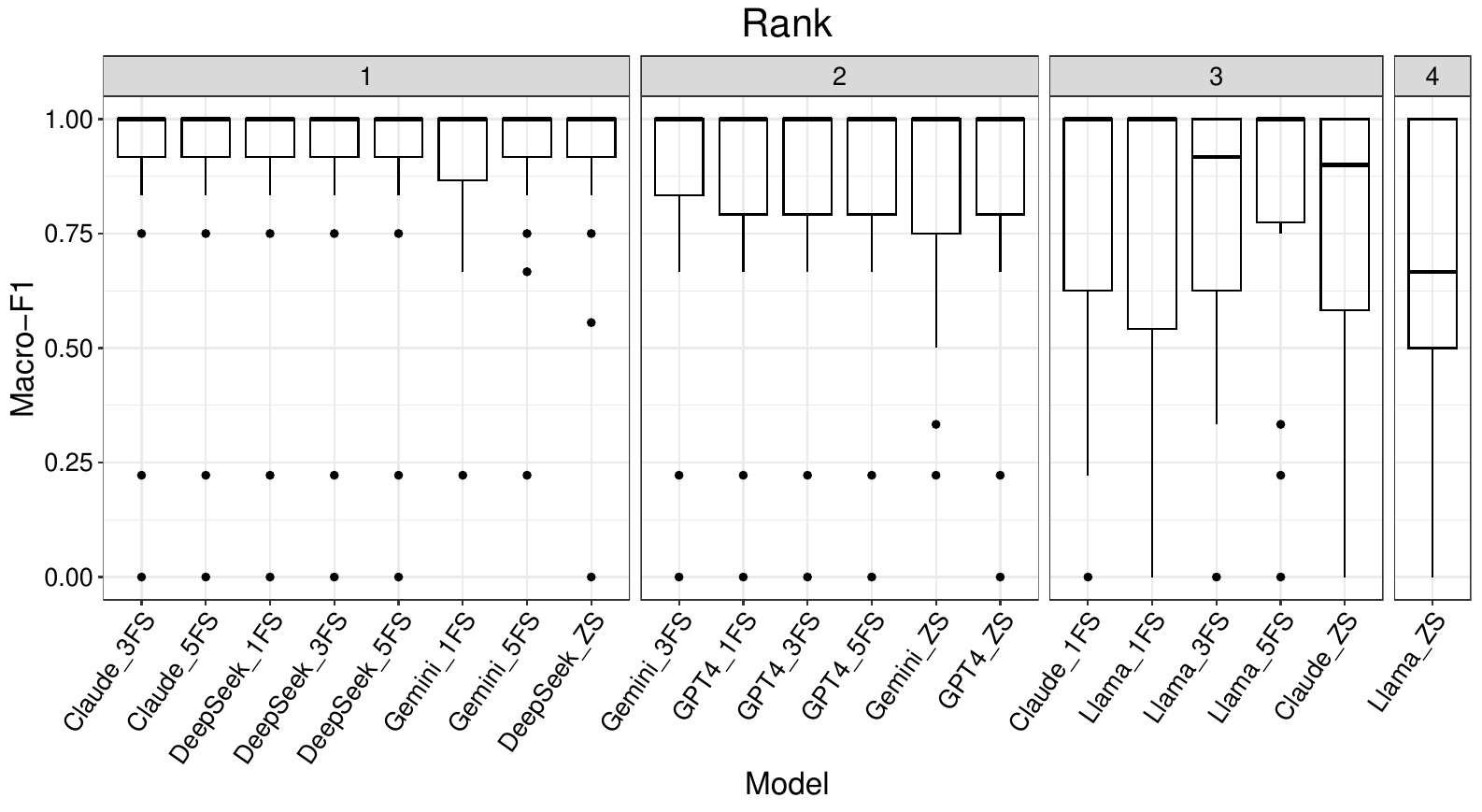}
\caption{Scott-Knott ESD ranking of zero-shot and few-shot promptings for each \ac{LLM} on the MC-\ac{NFR} task based on macro-F1 score}
 \label{fig:RQ2.2}
\end{figure}

In the MC-\ac{NFR} task, the results of the test, as presented in Figure \ref{fig:RQ2.2}, revealed that the top prefomers are: 
Claude\_3FS, Claude\_5FS, DeepSeek\_1FS, DeepSeek\_3FS, DeepSeek\_5FS, Gemini\_1FS, Gemini\_5FS, and DeepSeek\_ZS.
The second rank includes the following model-prompt configurations: Gemini\_3FS, GPT4\_1FS, GPT4\_3FS, GPT4\_5FS, Gemini\_ZS, and GPT4\_ZS. 

Notably, DeepSeek and GPT4 demonstrated consistent performance across zero-shot and few-shot configurations, with DeepSeek achieving the highest overall ranking, followed by GPT4. In contrast, Claude exhibited an improvement with three-shot and five-shot prompts, placing it among the top-performing models. Gemini also benefited from few-shot learning, with its performance under both one-shot and five-shot prompting ranking among the highest. Similarly, Llama showed improved outcomes when provided with few-shot examples.

\begin{figure} [hbtp]
 \center
 \includegraphics[width=.98
 \textwidth]{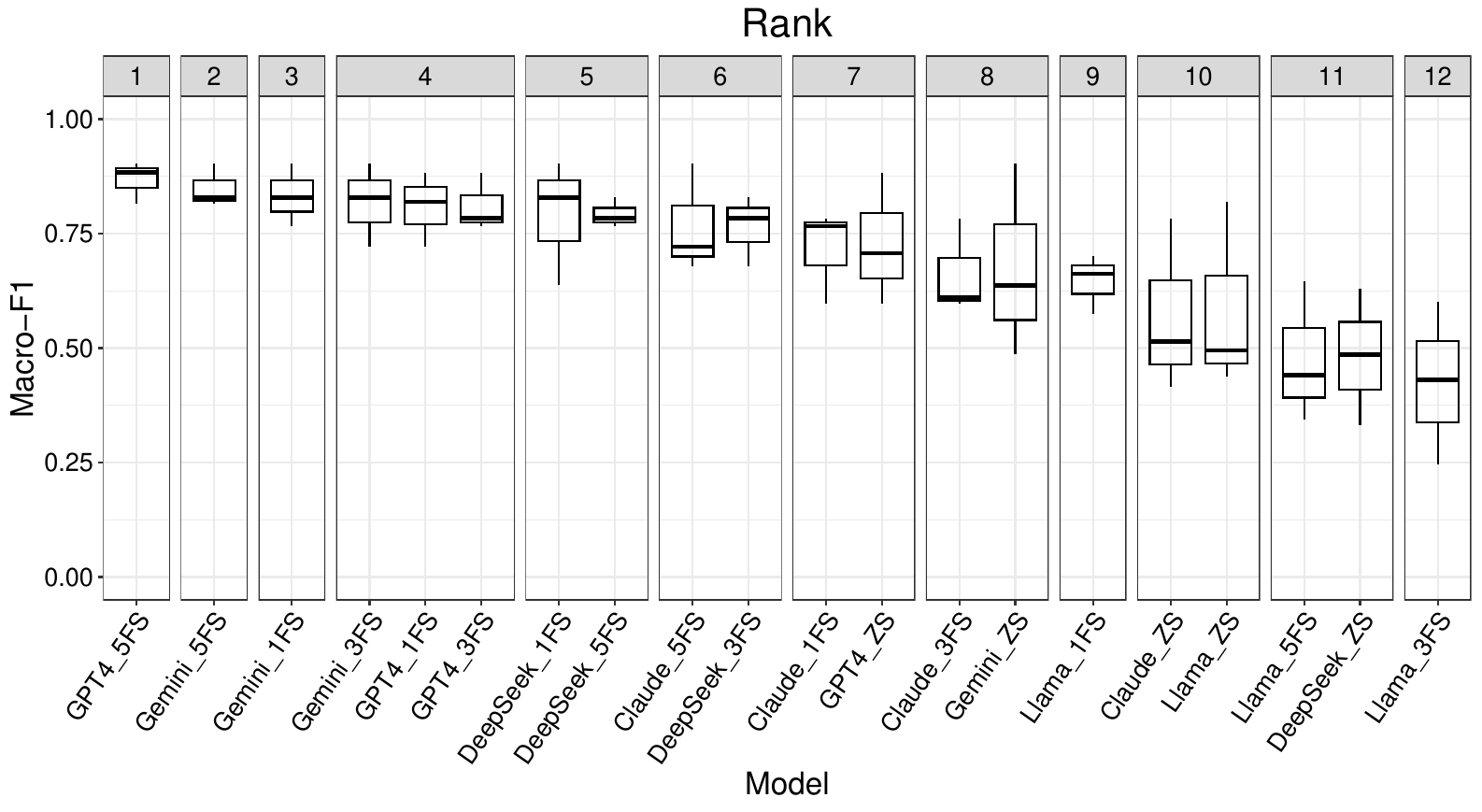}
\caption{Scott-Knott ESD ranking of zero-shot and few-shot promptings for each \ac{LLM} on the Sec-NonSec task based on macro-F1 score}
 \label{fig:RQ2.3}
\end{figure}

For Sec-NonSec task, the results presented in Figure \ref{fig:RQ2.3} indicate that GPT4-5FS achieved the highest performance, followed by Gemini-5FS. 

\subsection{RQ3: How do prompt framing strategies, specifically personas and \ac{CoT}, affect the classification performance of prompt-based \acp{LLM}?}

How do prompt framing strategies, specifically personas and \ac{CoT}, affect the classification performance of prompt-based \acp{LLM}?

The results of adding personas to the prompts are presented in Table \ref{tab:RQ3.1}, where an upward arrow ($\uparrow$) indicates an increase in the obtained macro-F1 score compared to its associated prompt with no persona, a downward arrow ($\downarrow$) indicates a decreased value, and an equals sign ($=$) indicates no change.

For the FR-\ac{NFR} task, the highest macro-F1 score was achieved by Gemini\_3FSP (0.92), followed by Gemini\_5FSP (0.90) and Claude\_3FSP (0.89). The lowest scores were recorded by Llama\_1FSP (0.25), Llama\_5FSP (0.37), and Llama\_3FSP (0.40). For the MC-\ac{NFR} classification task, DeepSeek\_3FSP achieved the highest score (0.94), followed by DeepSeek\_1FSP (0.90), while the lowest performances resulted from Llama\_ZSP (0.62), Llama\_1FSP (0.66), and Gemini\_ZSP (0.75). In the Sec-NonSec task, GPT4\_5FSP achieved the highest score (0.87), identical to its result without including personas, followed by GPT4\_3FSP with a macro-F1 score of 0.84. The lowest performers were Llama\_5FSP (0.13), Llama\_1FSP (0.15), and Llama\_3FSP (0.17). We note that the addition of persona resulted in a higher macro-F1 score for the MC-\ac{NFR} classification task, whereas for the other two tasks, the achieved scores were comparable to the highest scores obtained using few-shot prompts alone.

\begin{table}[htbp]
\caption{Macro-F1 scores of \acp{LLM} across software requirement classification tasks under zero-shot and few-shot prompting with persona}
\label{tab:RQ3.1}
\begin{tabular}{@{}llll@{}}
\toprule

 Model-prompt& FR-\ac{NFR}& MC-\ac{NFR}& Sec-NonSec \\\midrule
Claude\_ZSP & 0.82\textsuperscript{$\uparrow$} & 0.76\textsuperscript{$\uparrow$} & 0.7\textsuperscript{$\uparrow$} \\

Claude\_1FSP & 0.69\textsuperscript{$\uparrow$} & 0.82\textsuperscript{$\uparrow$} & 0.75\textsuperscript{$\uparrow$} \\

Claude\_3FSP & 0.89\textsuperscript{$\uparrow$} & 0.85\textsuperscript{$=$} & 0.72\textsuperscript{$\uparrow$} \\

Claude\_5FSP & 0.88\textsuperscript{$\uparrow$} & 0.85\textsuperscript{$=$} & 0.75\textsuperscript{$\downarrow$} \\
DeepSeek\_ZSP & 0.74\textsuperscript{$\uparrow$} & 0.88\textsuperscript{$=$} & 0.75\textsuperscript{$\uparrow$} \\
DeepSeek\_1FSP & 0.76\textsuperscript{$\uparrow$} & 0.9\textsuperscript{$\uparrow$} & 0.8\textsuperscript{$\uparrow$} \\
DeepSeek\_3FSP & 0.87\textsuperscript{$=$} & \textbf{0.94}\textsuperscript{$\uparrow$} & 0.75 \textsuperscript{$\downarrow$}\\
DeepSeek\_5FSP & 0.89\textsuperscript{$\uparrow$} & 0.89\textsuperscript{$\uparrow$} & 0.79\textsuperscript{$=$} \\
Gemini\_ZSP & 0.77\textsuperscript{$\uparrow$} & 0.75\textsuperscript{$\downarrow$} & 0.7\textsuperscript{$\uparrow$} \\
Gemini\_1FSP & 0.87\textsuperscript{$\uparrow$} & 0.84\textsuperscript{$\downarrow$} & 0.82\textsuperscript{$\downarrow$} \\
Gemini\_3FSP & \textbf{0.92}\textsuperscript{$\uparrow$} & 0.87\textsuperscript{$\uparrow$} & 0.76\textsuperscript{$\downarrow$} \\
Gemini\_5FSP & 0.9\textsuperscript{$\downarrow$} & 0.89\textsuperscript{$\downarrow$} & 0.83\textsuperscript{$\downarrow$} \\
GPT4\_ZSP & 0.52\textsuperscript{$\downarrow$} & 0.79\textsuperscript{$\downarrow$} & 0.79\textsuperscript{$\uparrow$} \\
GPT4\_1FSP & 0.69\textsuperscript{$\uparrow$} & 0.84\textsuperscript{$\uparrow$} & 0.82\textsuperscript{$\uparrow$} \\
GPT4\_3FSP & 0.77\textsuperscript{$\downarrow$} & 0.84\textsuperscript{$\uparrow$} & 0.84\textsuperscript{$\uparrow$} \\
GPT4\_5FSP & 0.81\textsuperscript{$\downarrow$} & 0.84\textsuperscript{$\uparrow$} & \textbf{0.87}\textsuperscript{$=$}\\
Llama\_ZSP & 0.79\textsuperscript{$\uparrow$} & 0.62\textsuperscript{$\downarrow$} & 0.6 \textsuperscript{$\uparrow$} \\
Llama\_1FSP & 0.25\textsuperscript{$\downarrow$} & 0.66\textsuperscript{$\downarrow$} & 0.15\textsuperscript{$\downarrow$} \\
Llama\_3FSP & 0.4\textsuperscript{$\downarrow$} & 0.89\textsuperscript{$\uparrow$} & 0.17\textsuperscript{$\downarrow$} \\
Llama\_5FSP & 0.37\textsuperscript{$\downarrow$} & 0.8\textsuperscript{$=$} & 0.13\textsuperscript{$\downarrow$} \\ \bottomrule
\end{tabular}
\end{table}

The results summarized in Table \ref{tab:RQ3.2} reveal distinct pattern across the three tasks when utilizing \ac{CoT} within the prompts, where an upward arrow ($\uparrow$) denotes an increase in the macro-F1 score relative to its prompt without adding \ac{CoT}, a downward arrow ($\downarrow$) indicates a decrease, and an equals sign ($=$) signifies no change.

In the FR-\ac{NFR} task, the highest macro-F1 score was achieved by both DeepSeek\_3FSCoT and Gemini\_3FSCoT (0.85), which is lower than its performance without CoT.
The lowest performer was Claude\_1FSCoT, with a macro-F1 score of 0.38. For the MC-\ac{NFR} classification task, DeepSeek\_5FSCoT achieved the highest score (0.87), which is lower than the highest score achieved through few-shot prompts.
In contrast, Llama\_3FSCoT preformed the worst, with a score of 0.56. In the Sec-NonSec, both Gemini\_1FSCoT and GPT4\_3FSCoT  achieved the top score (0.85), which is lower than the highest score achieved by few-shot only prompts. The weakest scores was recorded by Claude\_ZSCoT (0.5).

\begin{table}[htbp]
\caption{Macro-F1 scores of \acp{LLM} across software requirement classification tasks under zero-shot and few-shot prompting with \ac{CoT}}
\label{tab:RQ3.2}
\begin{tabular}{@{}llll@{}}
\toprule

 Model-prompt& FR-\ac{NFR}& MC-\ac{NFR}& Sec-NonSec \\\midrule
Claude\_ZSCoT & 0.5\textsuperscript{$\downarrow$} & 0.74 \textsuperscript{$\downarrow$}& 0.5 \textsuperscript{$\downarrow$}\\
Claude\_1FSCoT & 0.38\textsuperscript{$\downarrow$} & 0.77\textsuperscript{$\downarrow$} & 0.66 \textsuperscript{$\downarrow$}\\
Claude\_3FSCoT & 0.43\textsuperscript{$\downarrow$} & 0.77\textsuperscript{$\downarrow$} & 0.64\textsuperscript{$\downarrow$} \\
Claude\_5FSCoT & 0.54\textsuperscript{$\downarrow$} & 0.78\textsuperscript{$\downarrow$} & 0.67\textsuperscript{$\downarrow$} \\

DeepSeek\_ZSCoT & 0.79\textsuperscript{$\uparrow$} & 0.79 \textsuperscript{$\downarrow$}& 0.72\textsuperscript{$\uparrow$} \\

DeepSeek\_1FSCoT & 0.82\textsuperscript{$\uparrow$} & 0.85\textsuperscript{$=$} & 0.81\textsuperscript{$\uparrow$} \\

DeepSeek\_3FSCoT & \textbf{0.85}\textsuperscript{$\downarrow$} & 0.85 \textsuperscript{$=$}& 0.8\textsuperscript{$\uparrow$} \\

DeepSeek\_5FSCoT & 0.82 \textsuperscript{$\downarrow$}& \textbf{0.87} \textsuperscript{$\uparrow$}& 0.76\textsuperscript{$\downarrow$} \\

Gemini\_ZSCoT & 0.81\textsuperscript{$\uparrow$} & 0.76\textsuperscript{$\downarrow$} & 0.68\textsuperscript{$=$} \\

Gemini\_1FSCoT & 0.78\textsuperscript{$\downarrow$} & 0.84\textsuperscript{$\downarrow$} & \textbf{0.85}\textsuperscript{$\uparrow$} \\

Gemini\_3FSCoT & \textbf{0.85}\textsuperscript{$\downarrow$} & 0.77\textsuperscript{$\downarrow$} & 0.72\textsuperscript{$\downarrow$} \\

Gemini\_5FSCoT & 0.81\textsuperscript{$\downarrow$} & 0.78 \textsuperscript{$\downarrow$}& 0.7\textsuperscript{$\downarrow$} \\

GPT4\_ZSCoT & 0.57\textsuperscript{$\uparrow$} & 0.78 \textsuperscript{$\downarrow$}& 0.75\textsuperscript{$\uparrow$} \\

GPT4\_1FSCoT & 0.71\textsuperscript{$\uparrow$} & 0.76\textsuperscript{$\downarrow$} & 0.84 \textsuperscript{$\uparrow$}\\
GPT4\_3FSCoT & 0.69 \textsuperscript{$\downarrow$}& 0.76 \textsuperscript{$\downarrow$}& \textbf{0.85}\textsuperscript{$\uparrow$} \\

GPT4\_5FSCoT & 0.71\textsuperscript{$\downarrow$} & 0.77\textsuperscript{$\downarrow$} & 0.82\textsuperscript{$\downarrow$} \\

Llama\_ZSCoT & 0.66\textsuperscript{$\downarrow$} & 0.76\textsuperscript{$\uparrow$} & 0.71 \textsuperscript{$\uparrow$}\\

Llama\_1FSCoT & 0.5\textsuperscript{$\uparrow$} & 0.68\textsuperscript{$\downarrow$} & 0.67\textsuperscript{$\uparrow$} \\

Llama\_3FSCoT & 0.5\textsuperscript{$\downarrow$} & 0.56\textsuperscript{$\downarrow$} & 0.73\textsuperscript{$\uparrow$} \\

Llama\_5FSCoT & 0.57\textsuperscript{$\downarrow$} & 0.67\textsuperscript{$\downarrow$} & 0.77\textsuperscript{$\uparrow$}\\
\bottomrule
\end{tabular}
\end{table}

When augmenting the prompt with both persona and \ac{CoT}, model performance varied. Some models showed improvement, others experienced a decline, and some remained unchanged, as presented in Table \ref{tab:RQ3.3}. In the table, an upward arrow ($\uparrow$) indicates an increase in the macro-F1 score compared to the associated prompt without persona or \ac{CoT}, a downward arrow ($\downarrow$) indicates a decrease, and an equals sign ($=$) indicates no change. 

The highest macro-F1 score on the FR-\ac{NFR}  task was achieved by Gemini\_ZSPCoT with a score of 0.84. However, this is lower than the scores achieved by other previously examined prompt-model configurations on the same task. For the
MC-\ac{NFR} task, the highest macro-F1 score was 0.94, achieved by DeepSeek\_5FSPCoT. This score was also reached by another previously examined prompt-model configuration. In the Sec-NonSec task, GPT4\_3FSPCoT achieved the highest score of 0.87, which was also matched by another previously examined prompt-model configuration.

\begin{table}[htbp]
\caption{Macro-F1 scores of \acp{LLM} across software requirement classification tasks under zero-shot and few-shot prompting with peersona and \ac{CoT}}
\label{tab:RQ3.3}
\begin{tabular}{@{}llll@{}}
\toprule

 Model-prompt& FR-\ac{NFR}& MC-\ac{NFR}& Sec-NonSec \\\midrule
 Claude\_ZSPCoT & 0.39\textsuperscript{$\downarrow$} & 0.75\textsuperscript{$=$} & 0.56\textsuperscript{$\downarrow$} \\
Claude\_1FSPCoT & 0.37\textsuperscript{$\downarrow$} & 0.72\textsuperscript{$\downarrow$} & 0.77\textsuperscript{$\uparrow$} \\
Claude\_3FSPCoT & 0.43\textsuperscript{$\downarrow$} & 0.76\textsuperscript{$\downarrow$} & 0.73\textsuperscript{$\uparrow$} \\
Claude\_5FSPCoT & 0.5\textsuperscript{$\downarrow$} & 0.8\textsuperscript{$\downarrow$} & 0.69\textsuperscript{$\downarrow$} \\
DeepSeek\_ZSPCoT & 0.77\textsuperscript{$\uparrow$} & 0.81\textsuperscript{$\downarrow$} & 0.78 \textsuperscript{$\uparrow$}\\
DeepSeek\_1FSPCoT & 0.77\textsuperscript{$\uparrow$} & 0.86\textsuperscript{$\uparrow$} & 0.81\textsuperscript{$\uparrow$} \\
DeepSeek\_3FSPCoT & 0.84\textsuperscript{$\downarrow$} & 0.85\textsuperscript{$=$} & 0.8\textsuperscript{$\uparrow$} \\
DeepSeek\_5FSPCoT & 0.78 \textsuperscript{$\downarrow$}& \textbf{0.94}\textsuperscript{$\uparrow$} & 0.8\textsuperscript{$\uparrow$} \\
Gemini\_ZSPCoT & \textbf{0.84} \textsuperscript{$\uparrow$}& 0.73\textsuperscript{$\downarrow$} & 0.61\textsuperscript{$\downarrow$} \\
Gemini\_1FSPCoT & 0.8\textsuperscript{$\downarrow$} & 0.78\textsuperscript{$\downarrow$}  & 0.79\textsuperscript{$\downarrow$}  \\
Gemini\_3FSPCoT & 0.83\textsuperscript{$\downarrow$}  & 0.79\textsuperscript{$\downarrow$}  & 0.69\textsuperscript{$\downarrow$}  \\
Gemini\_5FSPCoT & 0.81\textsuperscript{$\downarrow$}  & 0.84\textsuperscript{$\downarrow$}  & 0.69\textsuperscript{$\downarrow$}  \\

GPT4\_ZSPCoT & 0.61\textsuperscript{$\uparrow$} & 0.78 \textsuperscript{$\downarrow$}& 0.79\textsuperscript{$\uparrow$} \\
GPT4\_1FSPCoT & 0.74\textsuperscript{$\uparrow$}  & 0.76\textsuperscript{$\downarrow$} & 0.84\textsuperscript{$\uparrow$}  \\
GPT4\_3FSPCoT & 0.71 \textsuperscript{$\downarrow$}& 0.76\textsuperscript{$\downarrow$} & \textbf{0.87}\textsuperscript{$\uparrow$}  \\

GPT4\_5FSPCoT & 0.72 \textsuperscript{$\downarrow$}& 0.77\textsuperscript{$\downarrow$} & 0.84 \textsuperscript{$\downarrow$}\\

Llama\_ZSPCoT & 0.67\textsuperscript{$=$} & 0.8\textsuperscript{$\uparrow$} & 0.77\textsuperscript{$\uparrow$} \\

Llama\_1FSPCoT & 0.47\textsuperscript{$\uparrow$}  & 0.83\textsuperscript{$\uparrow$}  & 0.82\textsuperscript{$\uparrow$}  \\
Llama\_3FSPCoT & 0.5 \textsuperscript{$\downarrow$} & 0.72\textsuperscript{$\downarrow$} & 0.85\textsuperscript{$\uparrow$}  \\
Llama\_5FSPCoT & 0.49\textsuperscript{$\downarrow$} & 0.87\textsuperscript{$\uparrow$}  & 0.78\textsuperscript{$\uparrow$}  \\

 \bottomrule

\end{tabular}
\end{table}

To understand which prompt-model configurations yielded the best results for each task, we applied the Scott-Knott test to all the examined prompt-model configurations analyzed in this study. We note that the number of reported top ranks varies across tasks, due to space limitation.

\begin{figure}
 \centering
 \includegraphics[width=.98
 \textwidth]{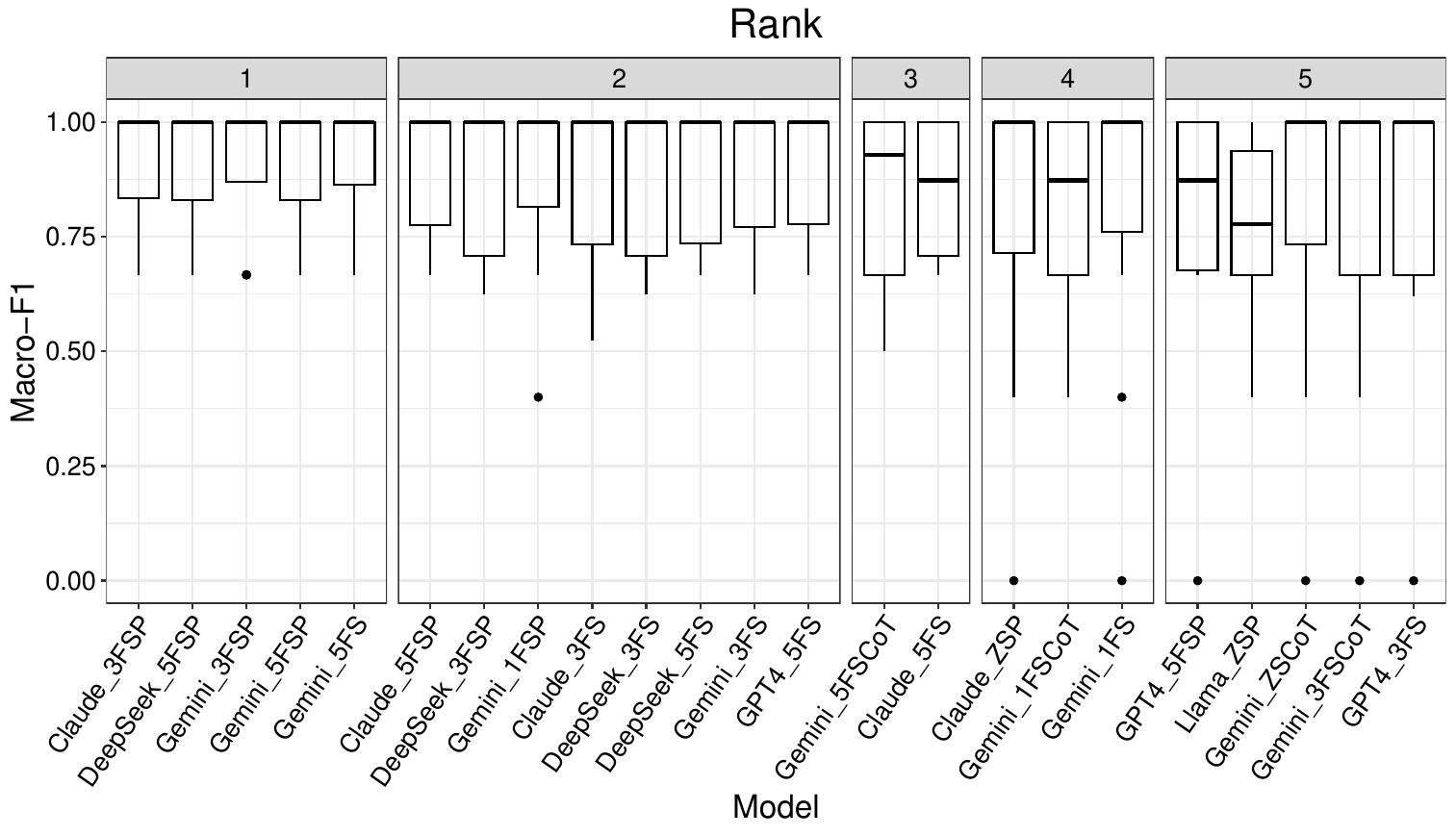}
\caption{Scott-Knott ESD ranking of the examined model-prompt configurations on the FR-\ac{NFR} task based on macro-F1 score}
 \label{fig:RQ3.1}
\end{figure}

As presented in Figure \ref{fig:RQ3.1}, for the FR-\ac{NFR} task, the top rank is largely dominated by few-shot-persona prompt-model configurations, with the exception of Gemini\_5FS, which was also ranked the best without a persona. The second rank is likewise dominated by few-shot prompts, with some configurations incorporating a persona and others not.

\begin{figure}
 \centering
 \includegraphics[width=.98
 \textwidth]{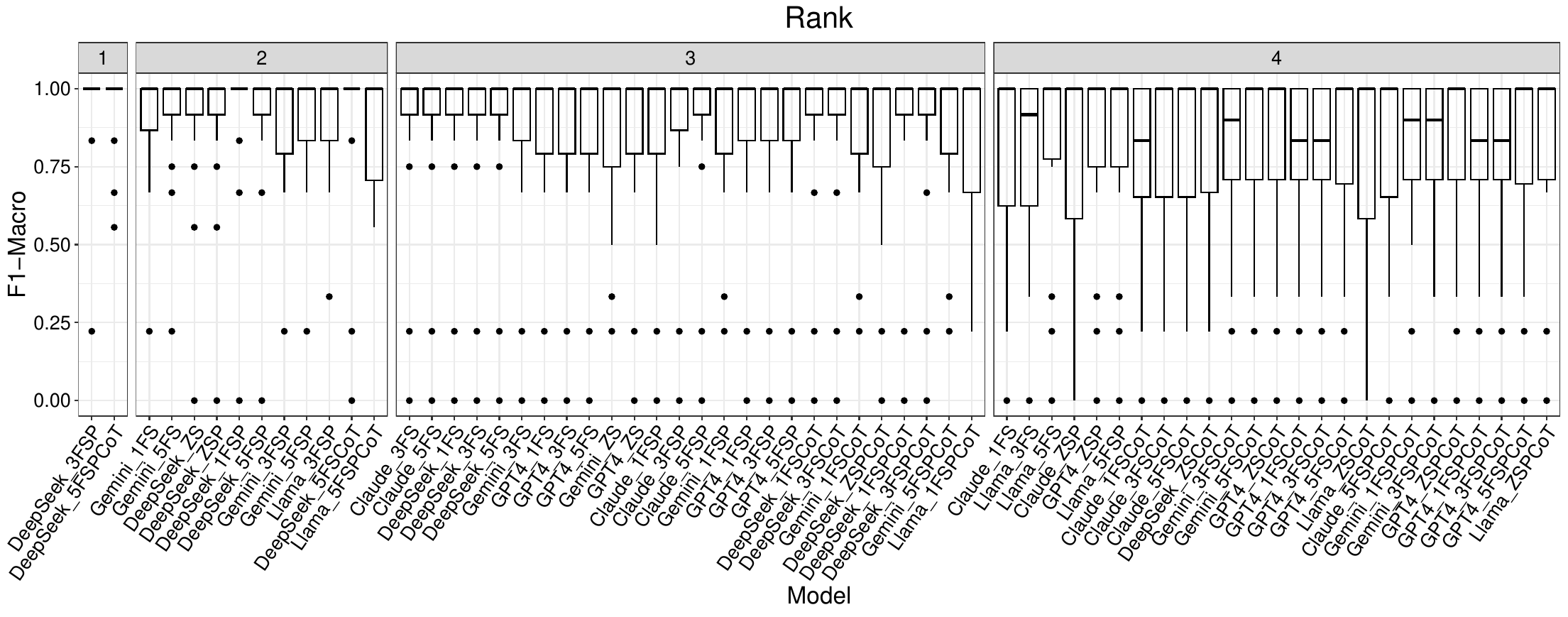}
 \caption{Scott-Knott ESD ranking of the examined model-prompt configurations on the MC-\ac{NFR} task based on macro-F1 score}
 \label{fig:RQ3.2}
\end{figure}

For the MC-NFR task, both DeepSeek\_3FSP and DeepSeek\_5FSPCoT were ranked the best, as presented in Figure \ref{fig:RQ3.2}. The second rank includes a variety of prompt-model configurations, including zero-shot, zero-shot with persona, few-shot, few-shot with persona, few-shot with \ac{CoT}, and few-shot with both persona and \ac{CoT}.

For the Sec-NonSec task, GPT4\_3FSPCoT, GPT4\_5FSP, and GPT4\_5FS were ranked highest. The second rank included various few-shot prompt–model configurations, some of which were augmented with \ac{CoT}, both persona and \ac{CoT}, or neither.
\begin{figure}
 \centering
 \includegraphics[width=.98
 \textwidth]{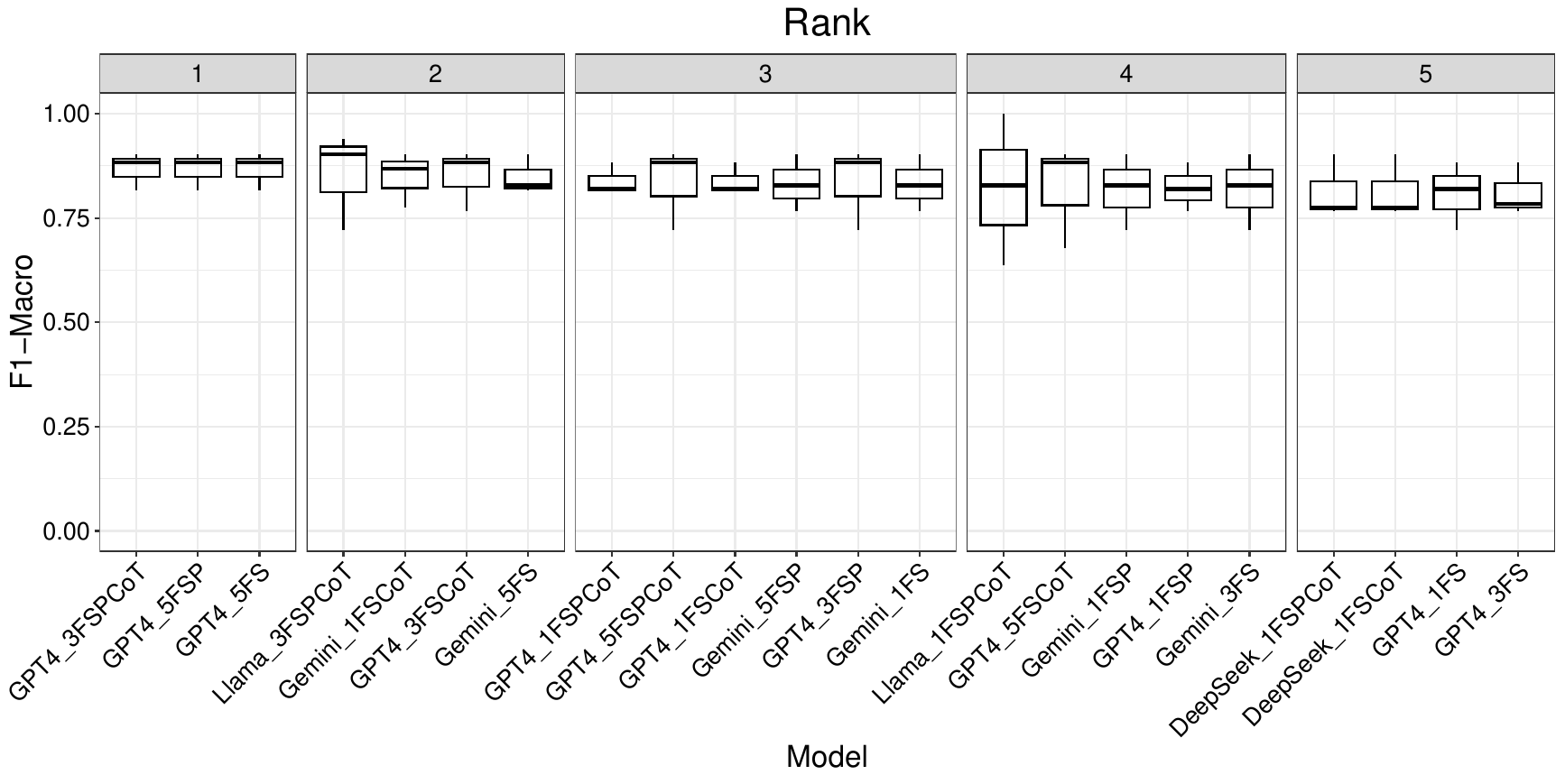}
 \caption{Scott-Knott ESD ranking of the examined model-prompt configurations on the Sec-NonSec task based on macro-F1 score}
 \label{fig:RQ3.3}
\end{figure}

\subsection{RQ4: How does the performance of prompt-based \acp{LLM} compare to the state-of-the-art fine-tuned transformer-based model for software requirements classification?}

The resulting macro-F1 scores of the best-performing prompt–\acp{LLM} configurations on each task and the fine-tuned transformer-based model (i.e., BERT) are presented in Table \ref{tab:RQ4}. Across all tasks, the highest obtained score was achieved by a prompt-based \ac{LLM}. On the FR-\ac{NFR} task, the majority of prompt-based \acp{LLM} performed better than BERT. On the MC-\ac{NFR} task, all prompt-based \acp{LLM} outperformed BERT, where half of these models outperformed BERT on the Sec-NonSec task.

\begin{table}[]
\caption{Macro-F1 scores of the best-performing prompts and \acp{LLM} configurations and the fine-tuned transformer-based models across requirement classification tasks }

\label{tab:RQ4}
\begin{tabular}{@{}llll@{}}
\toprule

Model-prompt or model& FR-\ac{NFR}& MC-\ac{NFR}& Sec-NonSec \\ \midrule
Claude\_3FSP & 0.89 & 0.85 & 0.72 \\
DeepSeek\_3FSP & 0.87 & \textbf{0.94} & 0.75 \\
DeepSeek\_5FSP & 0.89 & 0.89 & 0.79 \\
DeepSeek\_5FSPCoT & 0.78 & 0.94 & 0.8 \\
Gemini\_3FSP & \textbf{0.92} & 0.87 & 0.76 \\
Gemini\_5FS & 0.92 & 0.9 & 0.85 \\
Gemini\_5FSP & 0.9 & 0.89 & 0.83 \\
GPT4\_3FSPCoT & 0.71 & 0.76 & \textbf{0.87} \\
GPT4\_5FS & 0.88 & 0.82 & \textbf{0.87} \\
GPT4\_5FSP & 0.81 & 0.84 & \textbf{0.87}\\
BERT & 0.87 & 0.31 & 0.8 \\ \bottomrule
\end{tabular}
\end{table}

\begin{figure} [hbtp]
 \center
 \includegraphics[width=.85
 \textwidth]{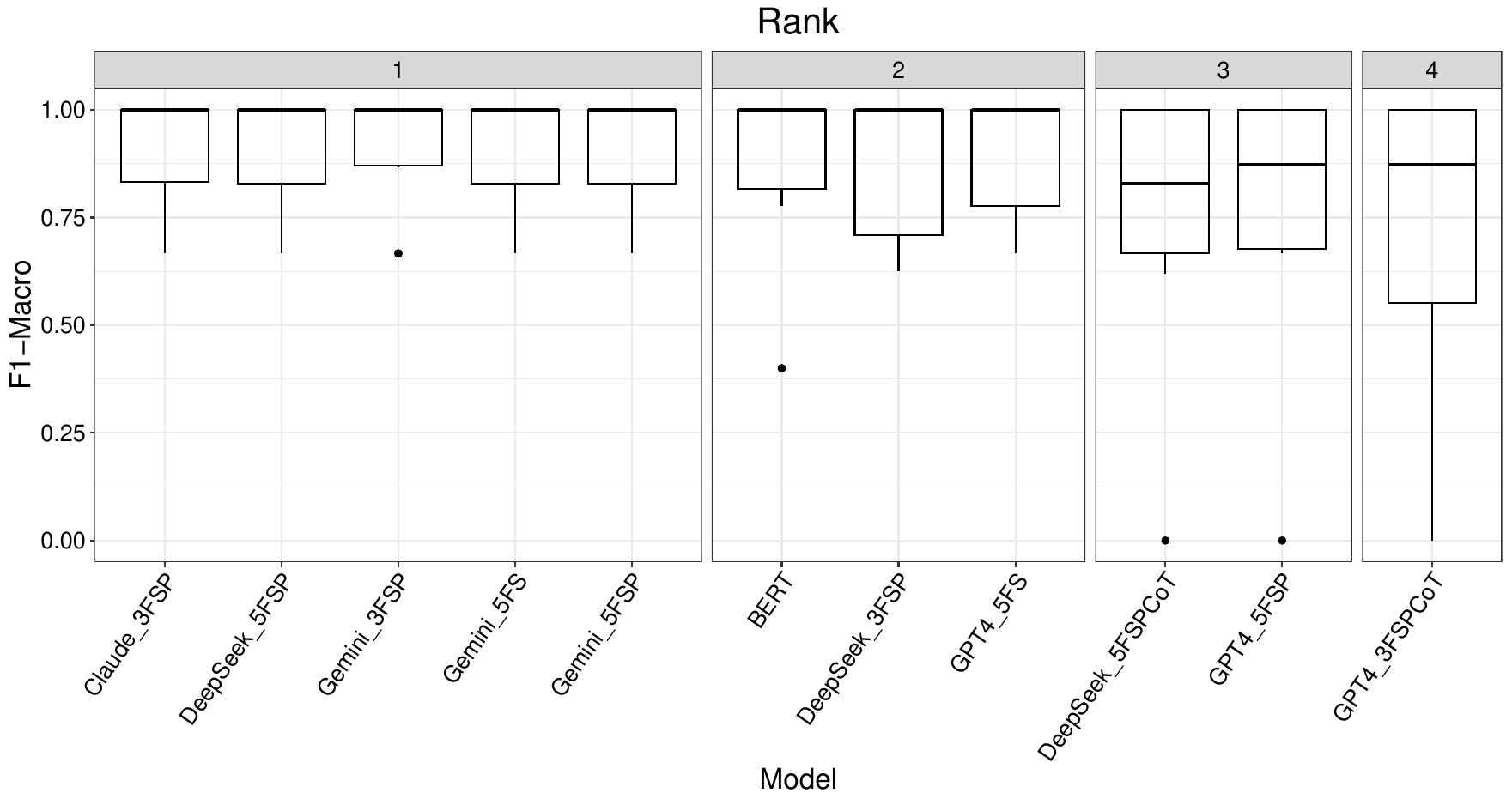}
\caption{Scott-Knott ESD ranking of the best-performing prompt-based \acp{LLM} and the fine-tuned transformer-based models on the FR-\ac{NFR} classification task based on macro-F1 score}
 \label{fig:RQ4.1}
\end{figure}

The results of the statistical test on the FR-\ac{NFR} task are presented in Figure \ref{fig:RQ4.1}. The test ranked half of prompt-based \acp{LLM} higher than BERT, two prompt-based \acp{LLM} were ranked similarly to BERT, while three models were ranked lower.

\begin{figure} [hbtp]
 \center
 \includegraphics[width=.85
 \textwidth]{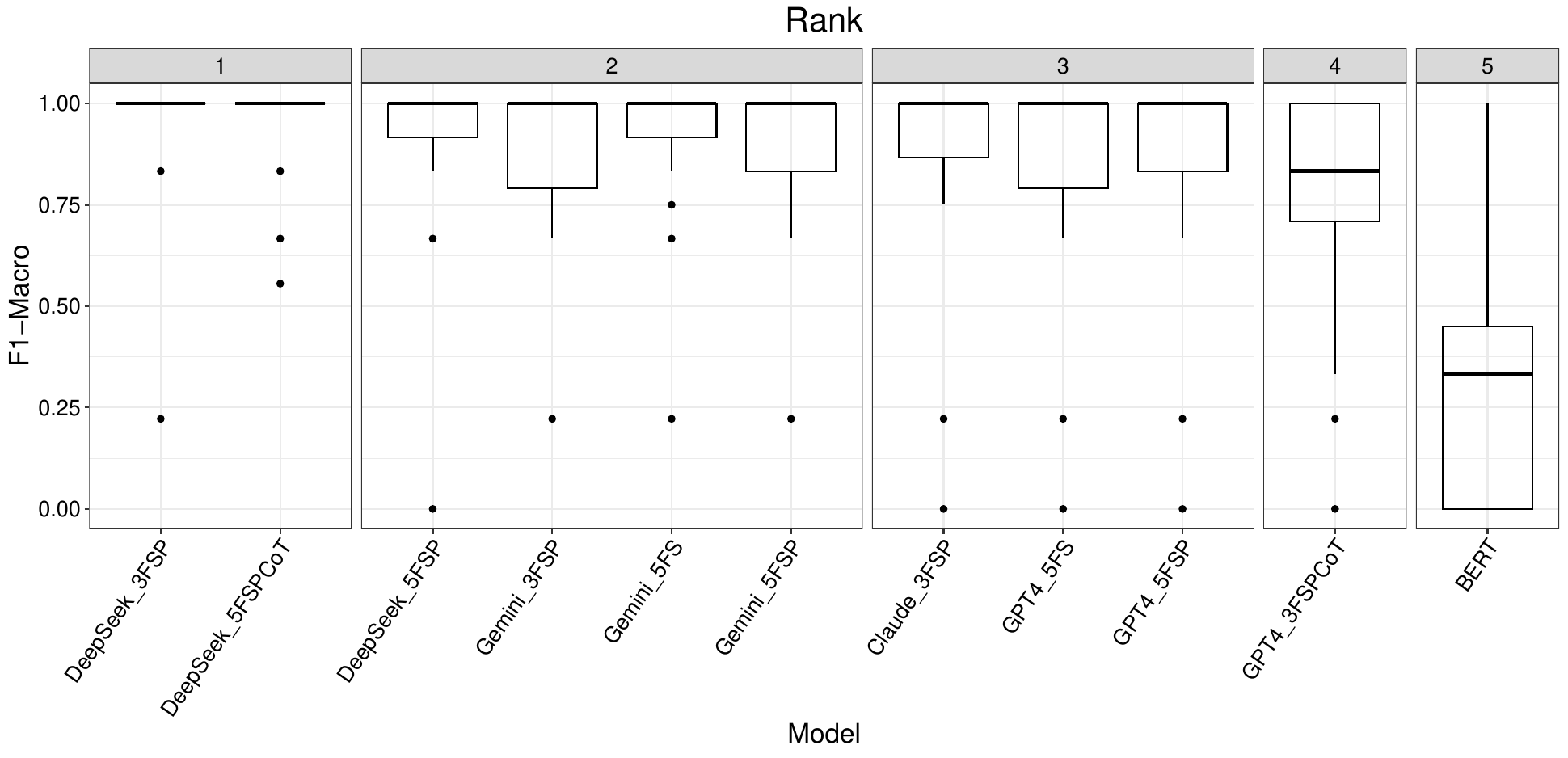}
\caption{Scott-Knott ESD ranking of the best-performing prompt-based \acp{LLM} and the fine-tuned transformer-based models on the MC-\ac{NFR} classification task based on macro-F1 score}
 \label{fig:RQ4.2}
\end{figure}

\begin{figure} [hbtp]
 \center
 \includegraphics[width=.85
 \textwidth]{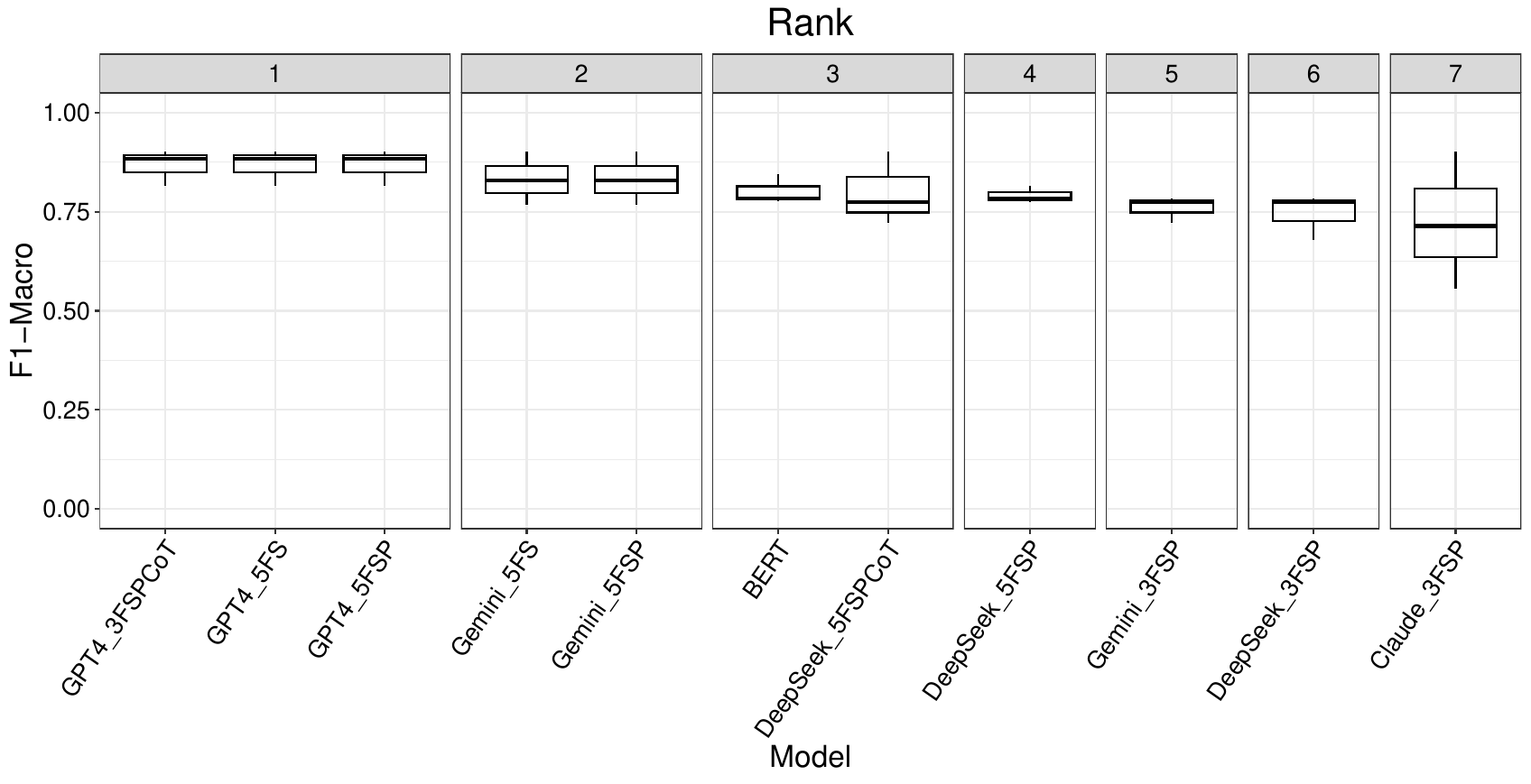}
\caption{Scott-Knott ESD ranking of the best-performing prompt-based \acp{LLM} and the fine-tuned transformer-based models on the Sec-NonSec classification task based on macro-F1 score}
 \label{fig:RQ4.3}
\end{figure}

On the MC-\ac{NFR} task, as presented in Figure \ref{fig:RQ4.2}, all prompt-based \acp{LLM} outperformed BERT. On the Sec-NonSec task, the results shown in Figure \ref{fig:RQ4.3} indicate that half of the prompt-based \acp{LLM} outperformed BERT, one model performed comparably to BERT, and the remaining four were ranked lower.

\section{Discussion} \label{Discussion}
The results for RQ1 demonstrate that no single prompt-based LLM consistently outperformed the others across all tasks under zero-shot prompting. Instead, performance varied depending on the specific task, with different models excelling in different areas. As evidenced by the statistical test results, Claude achieved the highest macro-F1 score on the FR–\ac{NFR} task. In contrast, GPT-4, DeepSeek, and Gemmini led on the MC–\ac{NFR} task, while GPT-4 outperformed all others in the Sec–NonSec classification. Notably, Gemmini consistently ranked either first or second across all tasks. This variation in results suggests that \acp{LLM} performance under zero-shot prompting is task-dependent, with no universal best performer. Therefore, one must carefully select models for each specific task, rather than assuming that a leading model in one context will generalize its superiority across others.

For RQ2, the models’ performance under few-shot configurations exhibited substantial variation. While some models benefited from few-shot prompting, others experienced performance declines, and a subset remained unaffected.
Although the few-shot rankings varied across tasks, indicating no single model consistently dominated, all three tasks consistently ranked Gemini\_5FS and GPT4\_5FS among the top performers, with their configurations placing first or second ranks. These findings suggest that despite the variability in performance introduced by few-shot learning, certain prompt-model configurations possess robust generalization and adaptability, positioning them as strong candidates for multi-task requirement classification.

As revealed by the results of RQ3, the incorporation of personas, \ac{CoT}, or both within prompts produced varying effects; some models benefited, others experienced a decline in performance, and some remained unaffected. Nevertheless, few-shot prompts consistently produced the highest-performing configurations. These findings suggest that few-shot prompting holds strong potential for software engineering requirement classification tasks. Moreover, augmenting few-shot prompts with either personas alone or a combination of persona and \ac{CoT} yielded the highest performance within the classification tasks, indicating that personas and combining personas with \ac{CoT} have the potential to enhance the effectiveness of prompt-based \acp{LLM} in software engineering requirement classification tasks.

Furthermore, although we were unable to draw a fair comparison between our work and previous studies due to several variations within the experimental settings, such as differences in the datasets used; evaluation metrics; and other, we compared our approach with the fine-tuned transformer-based model considered state of the art (i.e., BERT) in RQ4. The results exhibited a similar pattern: no prompt-model configuration could be identified as the best performer across all tasks. In all tasks, several prompt-model configurations outperformed BERT, revealing the potential of prompt-based LLMs in classifying requirements with minimal annotated data (i.e., the maximum we used was five examples). Although no model consistently performed the best, both Gemini\_5FS and Gemini\_5FSP were always ranked first or second across all tasks, surpassing BERT. This suggests their potential for utilization and generalization across software engineering requirement classification tasks.

While Gemini is a closed-source model, DeepSeek is an open-weight alternative that demonstrates competitive performance that is matching or even exceeding the performance of BERT on all tasks. DeepSeek can be freely downloaded and used for local deployment on supported hardware, without incurring API costs or restrictions.

As selecting a model for practical application requires consideration beyond empirical results, it is important to carefully balance performance with operational and long-term considerations. DeepSeek’s open-source nature provides advantages in transparency, accessibility, and privacy. In contrast, closed-source models may pose risks related to scalability, cost, limited transparency, and privacy.

Beyond the prompt-based \acp{LLM} empirical performance, the findings of this study carry several practical and theoretical implications. From a practical standpoint, the ability of prompt-based \acp{LLM}, particularly in few-shot configurations, to achieve performance on par with the state-of-the-art fine-tuned model suggests a more accessible and scalable pathway for organizations with limited annotated data or labeling resources. This is especially relevant for industrial settings, where the cost and time associated with retraining or fine-tuning domain-specific models can be prohibitive. Moreover, the use of persona and \ac{CoT} augmented prompts provides a flexible, low-code strategy for capturing domain nuances and improving interpretability without retraining or fine-tuning. These features position prompt-based \acp{LLM} as promising candidates for rapid application in real-world requirements engineering pipelines.

However, the flexibility of prompt-based \acp{LLM} comes with several practical considerations, including latency, cost, and privacy. As opposed to fine-tuned models that are optimized for specific tasks and deployed for fast inference, prompt-based \acp{LLM} may exhibit higher response times due to the need to interpret and process complex prompts at runtime. This latency could limit their suitability for time-sensitive applications or high-throughput systems. Moreover, the operational cost of commercial \acp{LLM} can become significant depending on model size, usage volume, and pricing policies. Another critical consideration is privacy, as using commercial \acp{LLM} may involve sending sensitive data to external servers and raises potential confidentiality risks.

While these issues remain theoretical within the scope of this study, as we did not empirically evaluate latency, privacy, or cost, they represent important considerations for practical deployment. Future work may systematically investigate these considerations across several prompting techniques, model sizes, and deployment environments to provide a more comprehensive understanding of their real-world implications.

Theoretically, these results challenge the long-held assumption that high-quality classification in requirements engineering necessitates large-scale supervised training. Instead, they underscore the viability of language models that rely on general language understanding, task framing, and contextual adaptation via prompts. This opens new avenues for research into prompt engineering as a substitute for task-specific fine-tuning, as well as the broader role of linguistic context and user intent modeling in software engineering tasks.

\section{Threats to validity} \label{Threats to validity} The validity of this study may be affected by threats to external, construct, internal, and conclusion validity. This section discusses these threats, along with the mitigation strategies adopted where applicable \cite{shull2007guide}.
\subsection{External validity}
A potential threat to the external validity of our study lies in the generalizability of its findings. Our results are based on two English-language requirements classification datasets and three classification tasks, and thus may not extend beyond the scope of these datasets and tasks. In particular, the performance of prompt-based \acp{LLM} may differ when applied to requirements written in other languages or to other types of requirements engineering tasks. Additionally, factors such as linguistic variation, input ambiguity, and domain-specific context may impact model performance, particularly in noisier or less structured settings.

Another potential external validity threat stems from the specific models and prompts employed in our study. As the findings are limited to the models and prompts evaluated, they may not extend to alternative configurations not considered in this study. We therefore acknowledge this as a limitation to the generalizability of our study.

Additionally, the utilization of closed-source prompt-based \acp{LLM} introduces further constraints on external validity. These models are subject to updates and version changes by their providers, which can lead to variability in model outputs over time. Such temporal fluctuations may affect the  consistency of results, potentially limiting the generalizability of our findings in future contexts.
\subsection{Construct validity}
One potential threat to the construct validity of this study concerns the choice of performance measures. To address this threat, we relied on macro-F1 as the primary measure for model comparisons, as it is well-suited for imbalanced datasets by ensuring equal weighting across all classes \cite{manning2008introduction}. This approach is consistent with that used in previous studies \cite{novielli2020can, zhang2023revisiting}.

Another potential threat to construct validity stems from the quality of dataset annotations. We rely on datasets annotated in prior studies that are adopted in previous, related work \cite{alhoshan2023zero,dekhtyar2017re}. Although efforts were made in the original work to ensure annotation accuracy, we cannot guarantee the complete correctness of these annotations. Therefore, any limitations in annotation quality are inherited from the source datasets.

An additional threat to construct validity in this study arises from the design of prompts. While we followed established guidelines for prompt construction, prompt formulation may still influence model behavior. The structure, wording, and contextual framing of prompts can affect how models interpret and respond to input, which may introduce bias. For example, although the inclusion of personas has the potential to enhance task alignment and clarity, it may also shape responses to reflect the persona rather than the underlying classification goal. Variations in prompt phrasing or assigned roles may also lead to inconsistent model behavior, raising concerns related to reliability, fairness, and inclusiveness. Therefore, we acknowledge this as a limitation of our study.

\subsection{Internal validity}
A potential threat to the internal validity of our study is the possibility of implementation errors. To mitigate this threat, we relied on widely accepted and validated implementations of the models and adhered to established guidelines and tutorials for their application in similar contexts, such as those provided by \cite{alammar2024hands, tunstall2022natural}.

\subsection{Conclusion validity}
Conclusion validity concerns the proper application and interpretation of statistical analyses \cite{shull2007guide}. We employed the non-parametric Scott-Knott ESD test to compare model performances. This test was chosen for its ability to form disjoint groups with meaningful effect sizes, its robustness to outliers, and its demonstrated effectiveness in prior studies within similar research contexts \cite{tantithamthavorn2018impact, puth2015effective}.

\section{Conclusion} \label{Conclusion}
Requirements classification in software engineering involves classifying natural language requirements into predefined classes, such as functional and non-functional requirements. Accurate classification is crucial for reducing development risks and enhancing the quality of software systems.

Several studies have leveraged advances in machine learning to automate requirements classification. While these efforts have yielded promising results, most of them rely on supervised learning approaches that depend heavily on large, annotated datasets for model training and validation. This reliance introduces several challenges: annotation is costly, time-consuming, often inconsistent, and requires substantial domain expertise. Moreover, the dependence on task-specific labeled data limits the adaptability of existing models, as new classification tasks typically require separate datasets and retraining or fine-tuning. This task-specific constraint further amplifies the challenge of dataset scarcity and remains a major barrier to developing automated requirements classification systems that are both effective and generalizable.

To address these challenges, this study explores the effectiveness of prompt-based \ac{LLM}s for requirement classification. We empirically evaluated and statistically compared the performance of the following five prompt-based \ac{LLM}s: Claude, DeepSeek, Gemini, GPT4, and Llama in combination with the following four prompting techniques: zero-shot, few-shot, persona, and \ac{CoT} prompting, across the following three requirements classification tasks: FR-NFR, where the goal is binary classification of functional and non-functional requirements; MC-NFR, where the goal is multi-class classification of non-functional requirements into their respective classes; and Sec-NonSec, where the goal is binary classification of security and non-security requirements. We then compared the performance of the highest-ranked prompt–model configurations on each task against that of the state-of-the-art fine-tuned transformer-based classification model (i.e., BERT) across all tasks.

The results of the first RQ revealed that under zero-shot prompting, no prompt-based \ac{LLM} consistently outperformed others across all tasks, as performance varied depending on the task. Moreover, the second RQ's results demonstrated that few-shot prompting led to improved classification performance across all tasks, and the results of the third RQ revealed that augmenting few-shot prompts with either persona alone or a combination of persona and \ac{CoT} yielded the best results across all tasks. Additionally, the final RQ’s comparison of the best-performing prompt-based \acp{LLM} for each task and BERT revealed that several prompt-based \acp{LLM} achieved higher performance across all tasks. Notably, the combination of five-shot prompting and Gemini, with and without persona augmentation, emerged as the overall best-performing prompt–model configuration.

The above results  underscores the potential of prompt-based \ac{LLM}s to advance requirements classification in software engineering. By reducing dependence on large annotated datasets and providing greater adaptability across tasks, prompt-based \ac{LLM}s represent a promising direction for enhancing intelligent requirements engineering tools and practices.

Future work can extend these findings by evaluating how different aspects of prompt design, such as alternative personas and few-shot example selection techniques, impact model performance. Further research may also broaden the investigation by considering computational efficiency and inference latency when comparing performance across both prompt-based \acp{LLM} and traditional supervised approaches. Additionally, detailed error analyses could reveal common failure patterns and guide the development of more effective and targeted prompt engineering practices for requirements classification.


\bibliography{sn-bibliography}

\end{document}